\documentclass[lettersize,journal]{IEEEtran}
\usepackage{amsmath,amsfonts}
\usepackage{algorithmic}
\usepackage{algorithm}
\usepackage{array}
\usepackage{amssymb}

\usepackage[caption=false,font=footnotesize,labelfont=rm,textfont=rm]{subfig}
\usepackage{titlesec}
\usepackage{color}
\usepackage{graphicx}
\usepackage{textcomp}
\usepackage{stfloats}
\usepackage{url}
\usepackage{verbatim}
\usepackage{graphicx}
\usepackage{epstopdf}
\usepackage{cite}
\begin{document}

\title{
	\huge{Affine Frequency Division Multiplexing Over Wideband Doubly-Dispersive Channels With Time-Scaling Effects}
}

\author{Xiangxiang Li, Haiyan Wang, \emph{Member, IEEE}, Yao Ge, \emph{Member, IEEE},  Xiaohong Shen, \emph{Member, IEEE}, \\ Yong Liang Guan,  \emph{Senior Member, IEEE}, Miaowen Wen,  \emph{Senior Member, IEEE}, and Chau Yuen, \emph{Fellow, IEEE}

	\thanks{This work was supported by the National Natural Science Foundations of China under Grant No.62031021.
	The work of Yao Ge was supported by the RIE2025 Industry Alignment Fund-Industry Collaboration Projects (IAF-ICP) Funding Initiative, as well as cash and in-kind contribution from the industry partner(s). The work of Chau Yuen was supported by National Research Foundation, Singapore and Infocomm Media Development Authority under its Future Communications Research \& Development Programme  FCP-NTU-RG-2024-025. (\emph{Corresponding authors: Haiyan Wang; Yao Ge}.)}
	
	\thanks{Xiangxiang Li, Haiyan Wang and Xiaohong Shen are with the School of Marine Science and Technology and the	Key Laboratory of Ocean Acoustics and Sensing, Ministry of Industry and Information Technology, Northwestern Polytechnical University, Xi’an, Shaanxi 710072, China. Haiyan Wang is also with the School of Electronic Information and Artifcial Intelligence, Shaanxi University of Science and Technology,	Xi’an, Shaanxi 710021, China (e-mail: lixx@mail.nwpu.edu.cn; hywang@nwpu.edu.cn; xhshen@nwpu.edu.cn).}
	\thanks{Yao Ge is with the Continental-NTU Corporate Laboratory, Nanyang Technological University, Singapore 639798 (e-mail: yao.ge@ntu.edu.sg).}
	\thanks{Yong Liang Guan and Chau Yuen are with the School of Electrical and Electronics Engineering, Nanyang Technological University, Singapore 639798 (e-mail: eylguan@ntu.edu.sg;	chau.yuen@ntu.edu.sg).}
	\thanks{Miaowen Wen is with the School of Electronic and Information Engineering, South China University of Technology, Guangzhou 510641, China (e-mail: eemwwen@scut.edu.cn).}
}

\maketitle
\setlength{\textfloatsep}{4.5pt}

\begin{abstract}
The recently proposed affine frequency division multiplexing (AFDM) modulation has been considered as a promising technology for narrowband doubly-dispersive channels. However, the time-scaling effects, i.e., pulse widening and pulse shortening phenomena, in extreme wideband doubly-dispersive channels have not been considered in the literatures.
In this paper, we investigate such wideband transmission and develop an efficient transmission structure with chirp-periodic prefix (CPP) and chirp-periodic suffix (CPS) for AFDM system. We derive the input-output relationship of AFDM system under 
time-scaled wideband doubly-dispersive channels
and demonstrate the sparsity in discrete affine Fourier (DAF) domain equivalent channels.  We further optimize the AFDM chirp parameters to accommodate the time-scaling characteristics in wideband doubly-dispersive channels and verify the superiority of the derived chirp parameters by pairwise error probability (PEP) analysis. We also develop an efficient cross domain distributed orthogonal approximate message passing (CD-D-OAMP) algorithm for AFDM symbol detection and analyze its corresponding state evolution.
By analyzing the detection complexity of CD-D-OAMP detector and evaluating the error performance of AFDM systems based on simulations, we demonstrate that the AFDM system with our optimized chirp parameters outperforms the existing competitive modulation schemes in time-scaled wideband doubly-dispersive channels.
Moreover, our proposed CD-D-OAMP detector can achieve the desirable trade-off between the complexity and performance, while supporting parallel computing to significantly reduce the computational latency.
\end{abstract}


\newpage
\section{Introduction}
\IEEEPARstart{T}{he} future wireless communication systems are expected to support the ultra-wideband communication for mobile scenarios, which motivates the development of an efficient scheme for high-reliable communications over wideband doubly-dispersive channels. 
Different from the Doppler effect in narrowband communication systems, the Doppler effect in wideband communication systems is characterized based on Doppler scale factor rather than simplified uniform Doppler frequency shift, which results in significant time-scaling effects and frequency-dependent non-uniform Doppler shifts \cite{d1,d2,d3}.

In narrowband doubly-dispersive channels, 
the widely used orthogonal frequency division multiplexing (OFDM) modulation maps the
transmitted symbols into time-frequency domain and achieves the high spectral efficiency. However, it suffers from significant performance loss due to the Doppler spread \cite{d4}, which disrupts the orthogonality between sub-carriers and results in the inter-carrier interference (ICI).
To address the limitations of OFDM, the orthogonal time frequency space (OTFS) \cite{d5, d6} modulation is proposed for high-mobility scenarios with large Doppler spread. By mapping the transmitted symbols into delay-Doppler domain rather than time-frequency domain, 
the OTFS can effectively utilize the delay-Doppler channel diversity for performance improvement. 
Although the orthogonal chirp division multiplexing (OCDM) \cite{d7} also shows better performance than OFDM by employing chirp signals as sub-carriers, it remains incapable to address the Doppler spread and the performance is limited to particular channel delay-Doppler profiles. For this issue, the 
recently proposed affine frequency division multiplexing
(AFDM) modulation \cite{d8,d9,d10,d11,d12,d13,d14,d141,d15} maps the transmitted symbols
into discrete affine Fourier (DAF) domain and optimize the chirp parameters based on channel spread characteristics,
which can effectively resolve channel paths and provide a complete delay-Doppler channel representation in DAF domain,
while achieving similar performance as OTFS. In addition, the AFDM also has lower modulation complexity and pilot guard overhead than OTFS \cite{d141}, while showing stronger advantage in joint channel, data and radar parameters estimation based on advanced parametric bilinear Gaussian belief propagation (PBiGaBP) and probabilistic data association (PDA) approaches compared to OTFS \cite{d15}, 
which has garnered significant attention.

%

However, compared to narrowband doubly-dispersive channels, the time-scaling effects, i.e., pulse widening and pulse shortening phenomena, in extreme wideband doubly-dispersive channels face great challenges, which result in a frequency-dependent non-uniform Doppler shift cross a relatively large spectrum band \cite{d1}. 
In such scenario, the orthogonality between OFDM sub-carriers is disrupted to varying degrees due to the frequency-dependent non-uniform Doppler shift, leading to an exacerbated performance degradation \cite{d16, d17}. In the OTFS system, the interference of delay-Doppler domain transmitted symbols increases significantly due to the influence  of non-uniform Doppler shift, while destroying the sparsity of the delay-Doppler domain equivalent channel \cite{d18, d19}. As a result, the OTFS fails to fully leverage the channel diversity for performance improvement.
The OCDM is unable to flexibly adjust the chirp rate for combating the impact of time-scaled wideband doubly-dispersive channels \cite{d20, d21}, resulting in a significant performance degradation. To address these limitations, the AFDM allows the flexible adjustment of chirp parameters for a full representation of delay-Doppler profile in the DAF domain \cite{d9}.
This capability motivates us to explore the design of AFDM systems and obtain the potential channel diversity gain for high-reliable communications over time-scaled wideband doubly-dispersive channels.

To achieve the potential performance gains promised by the AFDM system, efficient receivers are necessary for accurately recovering the original transmitted information.  
The classic linear minimum mean square error (LMMSE) detector \cite{d22, d23} suffers from significant performance loss and high computational burden due to the inversion of large-dimensional matrices.
The Gaussian message passing (GMP) \cite{d24, d25} and expectation propagation (EP) \cite{d26, d27, d28} detectors leverage the sparse channel structure for low-complexity detection, but they suffer significant performance loss due to short loops effect in dense factor graphs. The approximate message passing (AMP) \cite{d29} can achieve better detection under i.i.d. Gaussian channel matrices with linear complexity, but the real channel matrix does not satisfy this structure, leading to significant performance degradation \cite{d30}.
For general channel matrices, the orthogonal/vector AMP (OAMP/VAMP) \cite{d31, d32} can achieve more robustness performance by iteratively optimizing linear and nonlinear constraint modules. However, the LMMSE estimator is applied to linear optimization module in OAMP/VAMP for robust channel equalization, leading to high complexity due to large-dimensional matrix inversion. 
Although the matrix inversion operations can be approximated by various iterative process methods (such as Neumann Series \cite{d33}, Newton iteration \cite{d34}, Gauss-Seidel \cite{d35}, steepest descent \cite{d36}, conjugate gradient \cite{d37} and so on) to reduce complexity, there is still high processing latency due to a large number of iterations and the nature of the sequential updating. 
Note that the existing detectors mentioned above either suffer from high complexity and computational latency, or a significant performance loss. As the AFDM transmission over time-scaled wideband doubly-dispersive channels experience severe cross chirp sub-carrier interference, developing an efficient detector is important and challenging to guarantee the system performance.

In this paper, we examine the time-scaled wideband doubly-dispersive channels and develop an efficient transmission structure with the chirp-periodic prefix (CPP) and chirp-periodic suffix (CPS) for AFDM system to combat the delay and time-scaling effects in wideband mobile scenarios. We then re-optimize the AFDM chirp parameters to better accommodate time-scaling characteristics in 
wideband doubly-dispersive channels for a full representation of delay-Doppler profile in the DAF domain. To reduce the detection complexity and computational latency, 
we propose an efficient cross domain distributed OAMP (CD-D-OAMP) detector, which leverages the sparser time domain channel structure to trim the large-dimensional channel matrix into multiple manageable small-dimensional matrix groups, while ensuring the iterative convergence performance based on the unitary property of the cross domain transformation \cite{d38}.
The main contributions of this paper are summarized as follows:


%
\begin{itemize}
	\item{We develop a CPP and CPS transmission structure for AFDM system to combat the inter-symbol interference (ISI) under time-scaled wideband doubly-dispersive channels and derive the corresponding input-output relationship in the DAF domain.	
	We then demonstrate that the DAF domain equivalent channels of AFDM transmission still maintain sparsity under time-scaled wideband doubly-dispersive channels.}
	
	\item{We re-optimize the AFDM chirp parameters to prevent any resolvable delay or Doppler paths from overlapping in the DAF domain. Through pairwise error probability (PEP) theoretical analysis and maximum likelihood (ML) detection verification, we demonstrate that the performance of AFDM system with our optimized chirp parameters is superior to that of traditional narrowband chirp parameters and other existing modulation schemes in time-scaled wideband doubly-dispersive channels.}
	\item{We propose an efficient CD-D-OAMP iterative detector and analyze the corresponding state evolution, where the derived state evolution can closely match with the actual mean square error (MSE) performance and converge after a certain number of iterations.}
	\item{The simulation results demonstrate the superiority of AFDM systems with our optimized chirp parameters over the existing competitive modulation schemes in time-scaled wideband doubly-dispersive channels. Moreover, our proposed CD-D-OAMP detector can achieve a desirable trade-off between the complexity and performance, while significantly reducing the computational latency by parallel computing processing.
	}
\end{itemize}

The remainder of this paper is organized as follows: Section \uppercase\expandafter{\romannumeral2} introduces the AFDM system model and derives the input-output relationship in time-scaled wideband doubly-dispersive channels. In Section \uppercase\expandafter{\romannumeral3}, we optimize the AFDM chirp parameters according to time-scaling characteristics in
wideband doubly-dispersive channels. Section \uppercase\expandafter{\romannumeral4} analyzes the AFDM system performance based on PEP. In Section \uppercase\expandafter{\romannumeral5}, we propose the 
CD-D-OAMP detector and analyze the corresponding state evolution and detection complexity. Simulation results are provided in Section \uppercase\expandafter{\romannumeral6} and conclusion is drawn in Section \uppercase\expandafter{\romannumeral7}, respectively. 

\emph{Notations}: Bold capital letters denote matrices and bold lowercase letters denote vectors. $\mathbb{C}$ represents the complex numbers field. $\mathbb{E}$ represents the expectation. The superscripts $ \left( \cdot \right)^{\text{T}} $ and $ \left( \cdot \right)^{\text{H}} $ represent transpose and conjugate transpose, respectively. $ \mathbf{F}_N $ and $ \mathbf{F}_N^{\text{H}} $ denote the normalized $ N $-point discrete Fourier transform (DFT) matrix and inverse discrete Fourier transform (IDFT) matrix, respectively. The $ \mathbf{I}_{N} $ denotes the $ N \times N $ identity matrix. $ \text{diag} \left( \cdot \right)  $ denotes the diagonal matrix. The $ \boldsymbol{1}_{N} $  denotes $N$-point all-ones column vector. $[\cdot]_{N}$ denotes the mod $N$ operation. $ \mathcal {CN} \left( \cdot \right) $ denotes the complex Gaussian distribution. $\mathcal{O} \left( \cdot \right) $ denotes the computational complexity order. 

\section{AFDM System Model}
In this section, we firstly introduce the time-scaled wideband doubly-dispersive channel model and develop the CPP and CPS transmission frame structure for AFDM system. More importantly, we derive the input-output relationship of AFDM system and demonstrate that the DAF domain equivalent channels still retain sparsity in time-scaled wideband doubly-dispersive scenarios.

\subsection{Time-Scaled Wideband Doubly-Dispersive Channels}
In wideband communication systems (the transmission bandwidth $B$ is non-negligible relative to the carrier frequency $f_c$), we cannot simply approximate the Doppler effect as a uniform Doppler frequency shift relative to the carrier frequency, as is done in narrowband systems \cite{d9}. The wideband time-delay channel $g \left( t , \tau \right) $ can be described as \cite{d1}
\begin{align}\label{1}
	g \left( t , \tau \right) = \sum_{i=1}^{P} h_i  \delta [ \tau - \underbrace{\left( \tau_i - \alpha_i t \right)}_{\tau_i (t)} ]  e^{j 2 \pi \alpha_i f_c t},
\end{align}
where $P$ is the number of propagation paths. The parameters  ${h}_i$, $\tau_i \in [0,\tau_{max}]$ and $\alpha_i \in [-\alpha_{max} , \alpha_{max}]$ represent the complex gain, delay and Doppler scale factor of the $i$-th path, $\tau_{max} $ and $ \alpha_{max}$ denote the maximum delay and maximum Doppler scale factor, respectively. Note that the equivalent delay $\tau_i (t) = \tau_i - \alpha_i t$ is time-dependent, which results in a significant time-scaling effect.
Here, we denote the uniform Doppler frequency shift as $\nu_i = \alpha_i f_c$, which also corresponds to the Doppler effect in narrowband systems.
To further analyze the time-scaling characteristics in wideband systems, we transform the time-delay channel $g(t , \tau)$ into the time-frequency domain $G(t,f)$ as
\begin{align}\nonumber
	G(t,f) & = \int_{\tau} g(t,\tau) e^{-j 2 \pi f \tau} d \tau \\ \nonumber
	& = \sum_{i=1}^{P} h_i \int_{\tau} \delta [ \tau - \left( \tau_i - \alpha_i t \right) ]  e^{j 2 \pi \nu_i t}  e^{-j 2 \pi f \tau} d \tau   \label{2} \\ 
	& = \sum_{i=1}^{P} h_i e^{- j 2 \pi f \tau_i} e^{j 2 \pi (\nu_i + \alpha_i f) t}.
\end{align}

It is worth noting that the equivalent Doppler frequency shift $\nu_i (f) = \nu_i + \alpha_i f $ is non-uniform and frequency-dependent in wideband systems, which is actually different from the uniform Doppler shift $\nu_i$ observed in narrowband systems \cite{d9}. This characteristic is also know as Doppler squint effect (DSE) in wideband transmissions \cite{d18}.


\textbf{Remark 1}: \emph{To further illustrate the relationship between time-scaled wideband doubly-dispersive channels \cite{d1} considered in this paper and narrowband channels \cite{d9}, we consider a communication system with bandwidth $B$, carrier frequency $f_c$ and duration $T$, i.e. $t \in \left[ 0 , T\right] , f \in \left[ - \frac{B}{2} , \frac{B}{2} \right]  $. Then, the time-scaled wideband doubly-dispersive channels can be simplified to narrowband case with the following conditions \cite{d3}:}
	
\emph{(1) The transmission bandwidth $B$ is negligible relative to the carrier frequency $f_c$, i.e.} $B / f_c \ll 1$.

\emph{(2) The relative position of the transmitter and receiver does not change significantly relative to the positional resolution of the transmitted signal, i.e.} $ v / \tilde{c} = \alpha \ll 1 / (BT)$,\footnote{\label{n1}Note that the relative velocity and Doppler scale factor of $i$-th path are the corresponding $v_i$ and $\alpha_i$, respectively.} \emph{where} $v$ \emph{is the relative velocity of the transmitter and receiver,} $\tilde{c}$ \emph{is the propagation medium speed.}

\emph{Under the above two conditions, the wideband time-frequency response} $G(t, f)$ \emph{in \eqref{2}}  \emph{can be reduced to}
\begin{align}\label{3}
	G(t,f) \approx \sum_{i=1}^{P} {h}_i e^{- j 2 \pi f \tau_i} e^{j 2 \pi \nu_i  t },
\end{align}
\emph{which is consistent with the narrowband channel model \cite{d9}.}

\subsection{AFDM Transmission}
In this paper, we consider the AFDM transmission of $N$ constellation mapped symbols (such as QPSK, 16QAM) with duration $T = N \varDelta t$ and bandwidth $B = N \varDelta f$, where $\varDelta t = 1 / B$ and $\varDelta f = 1 / T$. The transmitted symbols $x[m], m= 0, 1, ... , N-1$ from a constellation alphabet $  \mathbb{A} \!=\! \left\lbrace a_1 , a_2 , ..., a_Q \right\rbrace $ are firstly mapped into the DAF domain.
Then, the time domain signal $x_{\text{T}} [n]$ is obtained by applying the inverse DAF (IDAF) transform to $x[m]$,
\begin{align}\label{4}
	x_{\text{T}} [n] = \frac{1}{\sqrt{N}} \sum_{m=0}^{N-1} x[m] e^{j 2 \pi \left(c_1 n^2 + m n / N + c_2 m^2 \right) },
\end{align}
where $n = 0, 1, ..., N-1$. $c_1$\footnote{\label{n2}Note that the AFDM chirp parameter $c_1 > 0$ or $c_1 < 0$ only represents the positive and negative chirp rates. In this paper, we primarily discuss the case of $c_1 > 0$ and the similar observations can be also obtained for the case of $c_1 < 0$.} and $c_2$ are the AFDM chirp parameters, which can be optimized to improve the system performance based on channel spread characteristics. We will discuss the detailed design guidelines of AFDM chirp parameters in Section \uppercase\expandafter{\romannumeral3}. 

To combat ISI and maintain the periodicity of AFDM symbols under time-scaled wideband doubly-dispersive channels, we develop an efficient CPP and CPS transmission frame structure for AFDM system. With the AFDM symbols periodicity defined in \cite{d9}, the CPP and CPS can be expressed as
\begin{subequations}\label{5}
	\begin{equation}
		{x}_{\text{T}} [n] = x_{\text{T}} [n + N] e^{- j 2 \pi c_1 \left( N^2 + 2 N n\right) } , n \in \left[ - L_{cpp} , 0 \right),
	\end{equation}
	\begin{equation}
		x_{\text{T}} [n] = x_{\text{T}} [n - N] e^{ j 2 \pi c_1 \left(  N^2 + 2 N n \right) } , n \in \left[  N , N + L_{cps}\right],
	\end{equation}
\end{subequations}
where $L_{cpp}$ and $L_{cps}$ are the length of added CPP and CPS, respectively. Here, we denote $T_{cpp} = L_{cpp} \varDelta t$ and $T_{cps} = L_{cps} \varDelta t$ as the continuous time length of CPP and CPS. For simplicity, we consider the AFDM symbol starting time of $1$-st path as the time reference, the correspondingly received observation window is given by $\left[ 0 , T \right) $. According to \eqref{1}, the time-scale $t$ of the $i$-th path AFDM symbol satisfies 
	\begin{align}\label{6}
		0 \leq t - \left( \tau_i - \alpha_i t \right) < T , i = 1, 2, ..., P.
\end{align}

\begin{figure}[t!]
	\centering
	\includegraphics[scale=0.45]{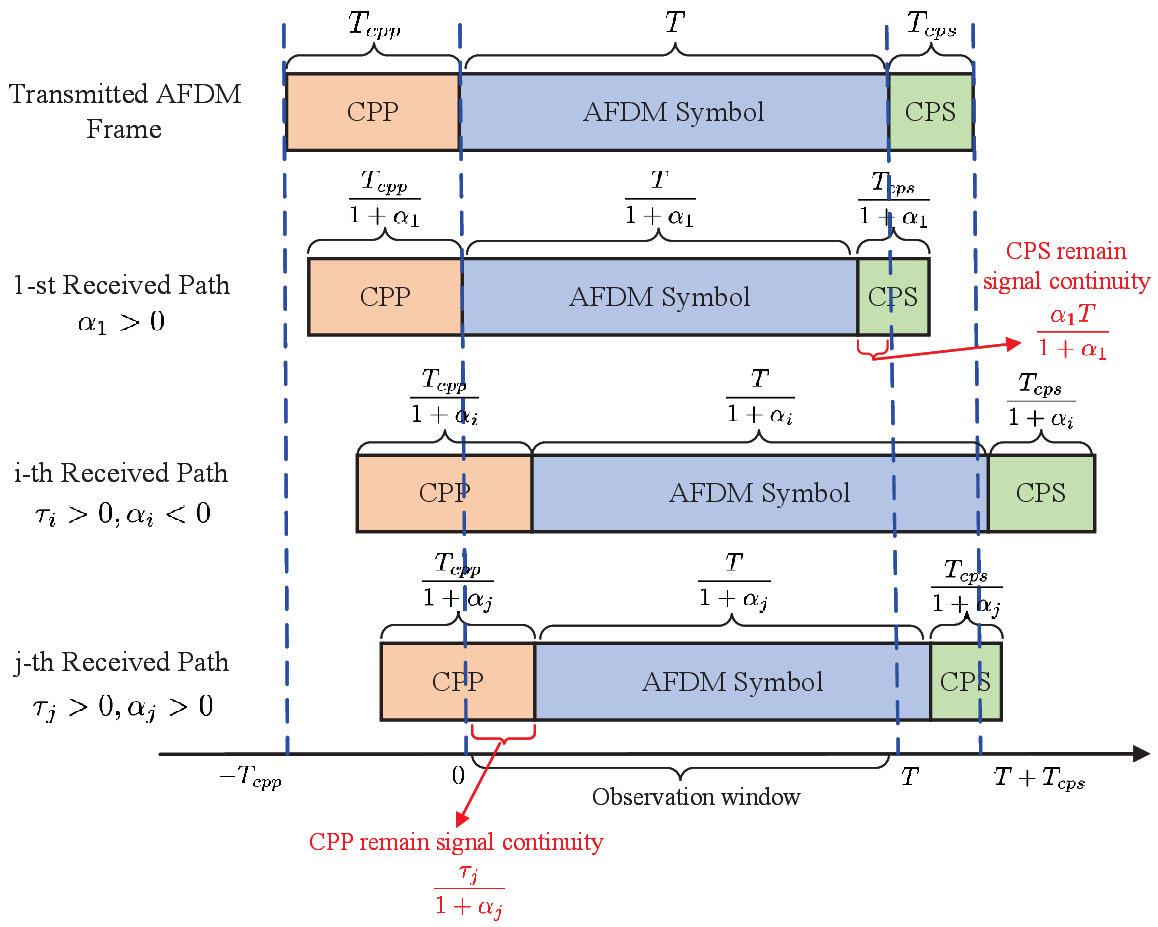}
	\caption{CPP and CPS transmission frame structure for AFDM system.}
	\label{fig_1}
\end{figure}

Consequently, the corresponding start and end time of the $i$-th path AFDM symbol are given by $\tau_i / (1 + \alpha_i)$ and $(T + \tau_i) / (1 + \alpha_i)$, as shown in Fig. \ref{fig_1}. Therefore, adding the CPP with duration $ T_{cpp} > \tau_{max} / (1 - \alpha_{max})$ and CPS with duration $ T_{cps} > \alpha_{max} T / (1 + \alpha_{max} )$ are essential to combat the delay and time-scaling effects and 
ensure the continuous periodicity of received AFDM symbols within the observation window.
The total duration $T_f$ of each AFDM frame is given by $T_f = T_{cpp} + T + T_{cps}$.\footnote{\label{n3}Note that the maximum Doppler scale factor $\alpha_{max}$ is typically small in practice. We can use the wideband underwater acoustic channels and the wideband wireless communication channels as examples for explanation. 
The maximum residual Doppler scale factor after resampling usually satisfies $\alpha_{max} < 10^{-4}$ in underwater acoustic channels \cite{d39, d40, d41} and $\alpha_{max} < 10^{-6}$ in wideband wireless radio communication channels \cite{d42, d43, d44}. Consequently, the duration of $T_{cps}$ is usually very small and has a negligible impact on spectral efficiency in practice.}

According to \cite{d10, d11, d12}, the correspondingly continuous time domain AFDM signal $x_{\text{T}}(t)$ can be given by 
\begin{align}\label{7}
	x_{\text{T}} (t) = \sum_{m=0}^{N-1} x[m] \Phi_{t} (m),  t \in \left[0 , T \right),
\end{align}
where 
\begin{align}\label{8}
	\Phi_{t} (m) = \frac{1}{\sqrt{N}} e^{j 2 \pi \left( c_1 \frac{N^2}{T^2} t^2 + \frac{m t}{T} + c_2 m^2 - \frac{\psi_{m, t} N}{T} t \right)   },
\end{align}
$\psi_{m, t}$ is the piece-wise function on intervals corresponding to the specific time partition, i.e.,
\begin{align}\label{9}
	\psi_{m, t} = 
	\begin{cases}
		0 , & 0 \leq t < \tilde{t}_{m,1} \\ 
		1 , & \tilde{t}_{m,1} \leq t < \tilde{t}_{m,2} \\
		~~\vdots, & ~~~~~~~\vdots \\
		\tilde{C}, & \tilde{t}_{m,\tilde{C}} \leq t < T
	\end{cases}
\end{align}
with
\begin{align}\label{10}
	\tilde{t}_{m, \rho} = 
	\begin{cases}
		0 , & \rho = 0 \\ 
		\left( \frac{N-m}{2N c_1} + \frac{\rho - 1}{2 c_1}\right) \frac{T}{N}  , & \rho = 1, 2, ..., \tilde{C}
	\end{cases}
\end{align}
and $\tilde{C} = 2 N c_1$.
We add the CPP and CPS to ${x}_{\text{T}} (t)$, then the  
time domain signal ${x}_{\text{T}}(t)$ is transmitted over the time-scaled wideband doubly-dispersive channel $ g(t, \tau)$. At the receiver, after removing the CPP and CPS, the received time domain signal $y_{\text{T}}(t) $ can be given by
\begin{align}\nonumber
	& y_{\text{T}}(t) = \int_{\tau} g(t,\tau) {x}_{\text{T}} \left( t - \tau \right) d \tau + \omega_{\text{T}} (t) \\ \nonumber
	&  = \sum_{i=1}^{P} h_i \int_{\tau} \delta [ \tau - \left( \tau_i - \alpha_i t \right) ]  e^{j 2 \pi \alpha_i f_c t} {x}_{\text{T}} \left( t - \tau \right) d \tau + \omega_{\text{T}} (t) \\ \label{11}
	 & = \sum_{i=1}^{P} h_i {x}_{\text{T}} \left(    t - (\tau_i - \alpha_i t) \right)   e^{j 2 \pi \nu_i t } + \omega_{\text{T}} (t),
\end{align}
where $\omega_{\text{T}} (t) $ is the additive white Gaussian noise (AWGN) with zero mean and variance $N_0$.

\textbf{Remark 2:}  \emph{Specially, if the AFDM system satisfies the narrowband approximation conditions mentioned in \textbf{Remark 1}, i.e.} $B = N \varDelta f \ll f_c $ \emph{and} $\alpha_i \ll 1 / (BT) = 1/N$, \emph{\eqref{11} can be simplified as}
\begin{align}\label{12}
	y_{\text{T}}(t) \approx \sum_{i=1}^{P} {h}_i {x}_{\text{T}} ( t - \tau_i )  e^{j 2 \pi \nu_i t } + \omega_{\text{T}} (t),
\end{align}
\emph{which is consistent with the narrowband system model \cite{d9}.}

Then, the discrete time domain received signal $ y_{\text{T}}[n] $ can be obtained by sampling $ y_{\text{T}}(t)$, 
\begin{align}\nonumber
	 & y_{\text{T}} [n] = y_{\text{T}}(t) \left. \right|_{t = n \varDelta t} \\ \label{13}
	& = \sum_{i = 1}^{P} {h}_i {x}_{\text{T}} \left(  (n - {\ell}_{i} + \alpha_i n) \varDelta t  \right)  e^{j 2 \pi  k_{i} n / N} + \omega_{\text{T}} (n \varDelta t) ,
\end{align}
where $n = 0, 1, ..., N-1$,  $ \ell_{i} = \tau_i / \varDelta t$ and  $  k_{i} = \nu_i N  \varDelta t$.

The corresponding input-output relationship of discrete time domain signal can be vectorized as
\begin{align}\label{14}
	\mathbf{y}_{\text{T}} = \bar{\mathbf{H}}_{\text{T}} \mathbf{x}_{\text{T}} + \boldsymbol{\omega}_{\text{T}} = \sum_{i=1}^{P} h_i \mathbf{H}_{\text{T},i} \mathbf{x}_{\text{T}} + \boldsymbol{\omega}_{\text{T}},
\end{align}
where $\bar{\mathbf{H}}_{\text{T}} = \sum_{i=1}^{P} h_i \mathbf{H}_{\text{T},i} $ is the time domain equivalent channel matrix and ${H}_{\text{T},i} [\bar{p},\bar{q}]$ is given by
\begin{align}\label{15}
	H_{\text{T},i}[\bar{p},\bar{q}] =  e^{j 2 \pi  \left[c_1 \left( \bar{p} - \ell_i + \alpha_i \bar{p}\right)^2 - c_1\bar{q}^2 + k_{i} \bar{p} / N \right]  } \mathcal{F}_{\text{T},i} [\bar{p},\bar{q}]
\end{align}
with
\begin{align}\nonumber
	& \mathcal{F}_{\text{T},i}  [\bar{p},\bar{q}] = \label{16} \\
	&\frac{1}{N} \sum_{m=0}^{N-1} e^{j 2 \pi \left\lbrace \left[ (1 + \alpha_i)\bar{p} \!-\! \bar{q} \!-\! \ell_i\right]  \frac{m}{N} \!-\! \left( \psi_{m, (\bar{p} \!-\! \ell_i \!+\! \alpha_i \bar{p}) \varDelta t } \!-\! \psi_{m, \bar{q}\varDelta t } \right) \right\rbrace}.
\end{align}

Finally, the AFDM output symbol $y[m]$ can be obtained by applying the DAF transform to $y_{\text{T}}[n]$,
\begin{align}\label{17}
	y[m] = \frac{1}{\sqrt{N}} \sum_{n=0}^{N-1} y_{\text{T}}[n] e^{-j 2 \pi \left(c_1 n^2 + m n / N + c_2 m^2 \right) } .
\end{align}

By substituting \eqref{7} and \eqref{13} into \eqref{17}, the input-output relationship of AFDM system in time-scaled wideband doubly-dispersive channels can be given by
\begin{align}\label{18}
	\mathbf{y} = \bar{\mathbf{H}} \mathbf{x} + \boldsymbol{\omega} = \sum_{i=1}^{P} {h}_i \mathbf{H}_i \mathbf{x} + \boldsymbol{\omega},
\end{align}
where $\boldsymbol{\omega}$ is the DAF domain noise with same distribution as $\boldsymbol{\omega}_{\text{T}}$. $\bar{\mathbf{H}} = \sum_{i=1}^{P} {h}_i \mathbf{H}_i $ is the DAF domain equivalent channel matrix and  $H_i \left[ p , q \right] $ is given by
\begin{align}\label{19}
	H_i[p,q] = \frac{1}{N} e^{j 2 \pi \left[ c_1 \ell_{i}^2 - q \ell_{i}/N + c_2 (q^2 - p^2) \right] } \mathcal{F}_i \left( p , q \right)
\end{align}
with
\begin{align}\nonumber
	&\mathcal{F}_i \left( p , q \right) \!=\! \label{20} \\
	&\sum_{n=0}^{N-1} \! e^{j 2 \pi \left\lbrace c_1 \left( \alpha_i^2 \!+\! 2 \alpha_i \right) n^2 \!-\! \left[ p \!-\! (1 \!+\! \alpha_i) q \!+\! 2 N c_1 (1 \!+\! \alpha_i) \ell_{i}  \!-\! k_{i} \!+\! \psi_{m, t_i'} \alpha_i N \right] \frac{n}{N} \right\rbrace }\!
\end{align}
and  $t_i' = (n - {\ell}_{i} + \alpha_i n) \varDelta t$. 
Alternatively, we can describe the relationship between time domain signal $\left\lbrace \mathbf{y}_{\text{T}}, \mathbf{x}_{\text{T}}, \boldsymbol{\omega}_{\text{T}} \right\rbrace $ and DAF domain signal $\left\lbrace \mathbf{y}, \mathbf{x}, \boldsymbol{\omega} \right\rbrace $ in vector representation with
\begin{align}\label{21}
	\mathbf{y} = \mathbf{U} \mathbf{y}_{\text{T}},   
	\mathbf{x} = \mathbf{U} \mathbf{x}_{\text{T}},  \boldsymbol{\omega} = \mathbf{U} \boldsymbol{\omega}_{\text{T}},
\end{align}
where $\mathbf{U} = {\boldsymbol{\Lambda}_{c_2} \mathbf{F}_{N} \boldsymbol{\Lambda}_{c_1}} $ is the DAF transform matrix, $\boldsymbol{\Lambda}_{c_b} = \text{diag} \{e^{-j2\pi c_b n^2}, n = 0, 1, ..., N-1, b = 1,2\}$, and the corresponding IDAF transform matrix is given by $\mathbf{U}^{\text{H}}$. 
Similarly, we can also describe the relationship between time domain equivalent channel $\bar{\mathbf{H}}_{\text{T}}$ and the DAF domain equivalent channel $\bar{\mathbf{H}}$ in matrix representation with
\begin{align}\label{22}
	\bar{\mathbf{H}} = \mathbf{U} \bar{\mathbf{H}}_{\text{T}} \mathbf{U}^{\text{H}}.
\end{align}

\subsection{Principle of Stationary Phase (POSP) Approximation}
To effectively explore the potential of AFDM systems in time-scaled wideband doubly-dispersive channels, it is essential to analyze the DAF domain equivalent channel structure $\bar{\mathbf{H}}$. From \eqref{19}, we observe that the structure of $H_i[p,q]$ primarily depends on $ \mathcal{F}_i \left( p , q \right) $. However, since  $ \mathcal{F}_i \left( p , q \right) $ involves summation of complex exponential terms with quadratic phase variation, deriving an analytical expression is challenging. 
To intuitively explain the structure of DAF domain equivalent channel, we employ the POSP method to obtain a simplified approximate analytical expression of $  \mathcal{F}_i \left( p , q \right) $. 

For the integral (summation for discrete signals) of a continuous complex exponential signal with quadratic phase variation and a slowly time-varying envelope, the POSP method can effectively approximate the true solution. A typical application of POSP method is in approximating the analytical solution of the linear frequency modulation (LFM) signal spectrum \cite{d45}.
To facilitate analysis, we can rewrite $\mathcal{F}_i \left( p , q \right)$ as
\begin{align}\label{23}
	\mathcal{F}_i \left( p , q \right) = \sum_{n=0}^{N-1} \underbrace{\text{Rect} \left( \frac{n}{N-1}\right) e^{j 2 \pi \theta_{p,q} \left( n \right) } }_{\mathcal{E}_i \left(p,q \right) } ,
\end{align}
where
\begin{subequations}\label{24}
	\begin{align}
		& \theta_{p,q} \left( n \right)  =  K n^2 / 2 - \varphi_{p,q} n, \\
		& K = 2 c_1 \left( \alpha_i^2 + 2 \alpha_i \right), \\
		& \varphi_{p,q} = \left[ p \!-\! (1 \!+\! \alpha_i) q \!+\! 2 N c_1 (1 \!+\! \alpha_i) \ell_{i}  \!-\! k_{i} \!+\! \psi_{m, t_i'} \alpha_i N \right]  / N.
	\end{align}
\end{subequations}

The $\text{Rect} \left( \cdot \right) $ denotes the rectangular window function, which is given by
\begin{align}\label{25}
	\text{Rect} \left( \frac{n}{N-1} \right) = 
	\begin{cases}
		1 , 0 \leq  \frac{n}{N-1} \leq 1 \\ 
		0 , \text{otherwise} 
	\end{cases}\!\!\!\!.
\end{align}

We can observe that the envelope of the signal $\mathcal{E}_i \left( p , q \right) $ remains constant over the integration interval, i.e., $\text{Rect} \left( \frac{n}{N-1}\right) $, which satisfies the POSP approximation constraints condition for a slowly time-varying envelope. According to POSP, the main contribution to the summation in \eqref{23} is from the nearby region of the stationary phase point. 
Then, the stationary phase point $\tilde{n}$ can be obtained by $\partial \theta_{p,q} / \partial n = 0$, i.e.,
\begin{align}\label{26}
	\tilde{n}  = \frac{\varphi_{p,q}}{K} 	= \frac{p \!-\! (1 \!+\! \alpha_i) q \!+\! 2 N c_1 (1 \!+\! \alpha_i) \ell_{i}  \!-\! k_{i} \!+\! \psi_{m, t_i'} \alpha_i N}{2 N c_1 \left( \alpha_i^2 \!+\! 2 \alpha_i \right) } .
\end{align}

According to the POSP method \cite{d45}, $  \mathcal{F}_i \left( p , q \right) $ can be finally approximated as
\begin{align}\label{27}
	\mathcal{F}_i \left( p , q \right) \approx \frac{1}{\sqrt{K}} \text{Rect} \left(  \frac{\tilde{n}}{N-1}   \right)  e^{j 2 \pi \left( \frac{K}{2} \tilde{n}^2 - \varphi_{p,q} \tilde{n}\right) } .
\end{align}

Here, we directly apply the conclusions of the POSP method and omit the mathematical derivations due to the space limitation. More details about POSP method can refer to \cite{d45}. Subsequently, the magnitude of $\left| H_i \left( p, q \right)  \right| $ can be approximated as
\begin{align}\nonumber
	\left| H_i \left( p, q \right)  \right| & = \left| \frac{1}{N} e^{j 2 \pi \left[ c_1 \ell_{i}^2 - q \ell_{i}/N + c_2 (q^2 - p^2) \right] } \mathcal{F}_i \left( p , q \right)  \right| \\ \label{28}
	& = \frac{1}{N} \left| \mathcal{F}_i \left( p , q \right) \right| \approx \frac{1}{N \sqrt{K}} \text{Rect} \left(  \frac{\tilde{n}}{N-1}   \right).
\end{align}

\subsection{Equivalent Channel Sparsity Analysis}
According to \eqref{28}, the $i$-th path non-zero elements satisfy
\begin{align}\label{29}
	0 \leq \frac{p - (1 + \alpha_i) q + 2 N c_1 (1 + \alpha_i) \ell_{i}  - k_{i} + \psi_{m, t_i'} \alpha_i N }{2 N (N-1) c_1 \left( \alpha_i^2 + 2 \alpha_i \right) } \leq 1.
\end{align}

It is worth noting that we primarily discuss the case of 
$c_1 > 0$ in this paper, and the maximum Doppler scale factor usually satisfies $\alpha_{max} < 10^{-4}$, i.e., $ |2 \alpha_{i}| > \alpha_{i}^2 $. Therefore, it can be shown that the $p$-th row of DAF domain equivalent channel in the $i$-th path satisfies $ \left|  H_i \left( p , q \right) \right| \neq 0 $ under POSP approximation only when
\begin{align}\label{30}
	q \in [Q_{\ell_i,\alpha_i} , \tilde{Q}_{\ell_i,\alpha_i}]_N,
\end{align}
where 
\begin{subequations}\label{31}
\begin{align}
	Q_{\ell_i,\alpha_i} \!\!=\! \!
	\begin{cases}
		2 N c_1 \ell_{i} \!+\! \frac{p  \!-\! k_{i} \!-\! 2  c_1 \left( \alpha_i^2 \!+\! 2 \alpha_i \right) N \left( N \!-\! 1 \right) + \psi_{m, t_i'} \alpha_i N }{1 \!+\! \alpha_i},\!\!\!\!\! & \alpha_i \! \geq  \! 0 \\ 
		2 N c_1 \ell_{i} + \frac{p  - k_{i} + \psi_{m, t_i'} \alpha_i N}{1 + \alpha_i} \!,\! &  \alpha_i \!<\! 0
	\end{cases}
\end{align}
and
\begin{align}
	\tilde{Q}_{\ell_i,\alpha_i} \!\!=\! \!
	\begin{cases}
		2 N c_1 \ell_{i} + \frac{p  - k_{i} +\psi_{m, t_i'} \alpha_i N}{1 + \alpha_i}, & \alpha_i \! \geq 0  \! \\ 
		2 N c_1 \ell_{i} \! + \! \frac{p  \!-\! k_{i} \!-\! 2  c_1 \left( \alpha_i^2 \!+\! 2 \alpha_i \right) N \left( N \!-\! 1 \right)  + \psi_{m, t_i'} \alpha_i N}{1 \!+\! \alpha_i},\!\!\!\!\! & \alpha_i \!< \! 0
	\end{cases}
	\!.\!
\end{align}
\end{subequations}

According to \eqref{9}, $\psi_{m, t_i'}$ is the piece-wise function and $\psi_{m, t_i'} \in [0, 2 N c_1]$, we then further strengthen the boundary in \eqref{31} as
\begin{subequations}\label{32}
	\begin{align}
		Q_{\ell_i,\alpha_i} \!=\! 
		\begin{cases}
			2 N c_1 \ell_{i} \!+\! \frac{p  \!-\! k_{i} \!-\! 2  c_1 \left( \alpha_i^2 \!+\! 2 \alpha_i \right) N \left( N \!-\! 1 \right)}{1 \!+\! \alpha_i},\! & \alpha_i  \geq 0 \\ 
			2 N c_1 \ell_{i} + \frac{p  - k_{i} + 2 c_1 \alpha_i N^2 }{1 + \alpha_i} \!,\! &  \alpha_i < 0
		\end{cases}
	\end{align}
	and
	\begin{align}
		\tilde{Q}_{\ell_i,\alpha_i} \! =\! 
		\begin{cases}
			2 N c_1 \ell_{i} + \frac{p  - k_{i} + 2 c_1 \alpha_i N^2}{1 + \alpha_i}, & \alpha_i  \geq 0 \\ 
			2 N c_1 \ell_{i} \! + \! \frac{p  \!-\! k_{i} \!-\! 2  c_1 \left( \alpha_i^2 \!+\! 2 \alpha_i \right) N \left( N \!-\! 1 \right) }{1 \!+\! \alpha_i}\!,\! & \alpha_i < 0
		\end{cases}
		\!.\!
	\end{align}
\end{subequations}

\begin{figure}[t!]
	\centering
	\includegraphics[scale=0.55]{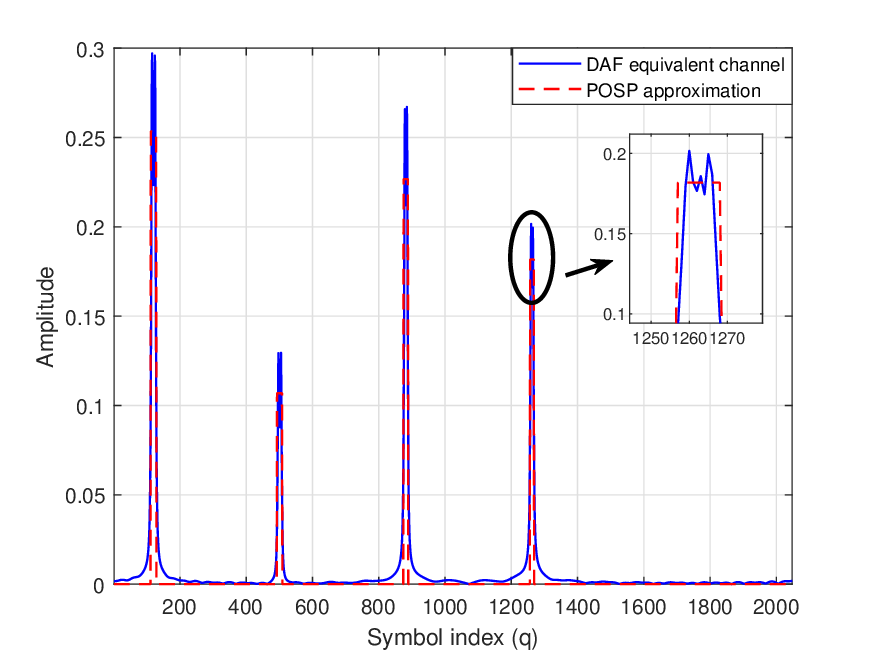}
	\caption{Comparison between DAF domain equivalent channel with POSP approximation.}
	\label{fig_2}
\end{figure} 

\begin{figure*}[hb]
	\hrulefill
	\begin{align}\label{33}
		H_i \left( p , q \right) \approx 
		\begin{cases}
			\frac{1}{N} e^{j 2 \pi \left[ c_1 \ell_{i}^2 - q \ell_{i}/N + c_2 (q^2 - p^2) \right] } \mathcal{F}_i \left( p , q \right) , q \in [ Q_{\ell_i , \alpha_i} - N_v ,  \tilde{Q}_{\ell_i , \alpha_i} + N_v ]_N \\ 
			0 , \text{otherwise} 
		\end{cases}\!\!\!\!.
	\end{align}
\end{figure*}

Fig. \ref{fig_2} compares the actual DAF domain equivalent channel matrix $|\bar{\mathbf{H}}|$ with POSP approximation in a specific $p$-th row. 
Without loss of generality, we take the time-scaled wideband underwater acoustic channel as an example to illustrate the equivalent channel structure,
where the maximum Doppler scale factor is $\alpha_{max} = 10^{-4}$ (after resampling processing \cite{d39,d40,d41}). Specifically, we consider four paths with $P = 4$ and set $N = 2048$, $p = 128$. The carrier frequency is $f_c = 6$kHz with sub-carriers spacing $\varDelta f = 4$Hz \cite{d41}. From Fig. \ref{fig_2}, we observe that
the non-zero regions of all paths based on POSP approximation can effectively match the peak regions of the actual DAF domain equivalent channel.
As $q$ moves away from the  non-zero elements region $[Q_{\ell_i,\alpha_i} , \tilde{Q}_{\ell_i,\alpha_i}]_N$, the magnitude of actual DAF domain equivalent channel rapidly decreases. Therefore, we only consider the interval $[ Q_{\ell_i,\alpha_i} - N_v ,  \tilde{Q}_{\ell_i,\alpha_i} + N_v]_N  $ as the dominant non-zero elements and ignore other small amplitude regions, which form an obvious sparse structure. 
$N_v$ is chosen such that $\left| H_i \left( p, {q} \right)  \right| $ holds the significant magnitude when $q \in [ Q_{\ell_i,\alpha_i} - N_v ,  \tilde{Q}_{\ell_i,\alpha_i} + N_v]_N  $. Meanwhile, the contributions from $q$ outside this interval are negligible and can be disregarded. Then, the sparse  representation of $i$-th path $\left| H_i \left( p, q \right)  \right| $ is written in \eqref{33}, shown at the bottom of this page.
From \eqref{33}, the width $\mathcal{L}_i$ of $i$-th path non-zero elements can be given by
\begin{align}\nonumber
	 \mathcal{L}_i & = \tilde{Q}_{\ell_i , \alpha_i} -   {Q}_{\ell_i , \alpha_i} +   2 N_v + 1  \\ \label{34}
	 & = \left| \frac{2  c_1 \left( \alpha_i^2 + 2 \alpha_i \right) N \left( N-1 \right) + 2 c_1 \alpha_i N^2 }{1 + \alpha_i} \right|  + 2 N_v + 1.
\end{align}

Note that the $\mathcal{L}_i$ is proportional to AFDM chirp parameter $c_1$, the number of transmitted symbols $N$ and the Doppler factor $\alpha_i$. It will reduce the sparsity of DAF domain equivalent channels under severe Doppler spread or long data transmission scenarios. 
Therefore, developing an efficient AFDM symbol detection algorithm in time-scaled wideband doubly-dispersive channels becomes a more important and challenging problem to exploit such underlying channel diversity. Finally, the input-output relationship of \eqref{18} can be expressed as 
\begin{align}\nonumber
	y \left[ p \right] \approx & \sum_{i=1}^{P} \frac{1}{N} {h}_i  \sum_{q = [ Q_{\ell_i,\alpha_{i}} \!-\! N_v ]_N }^{[ \tilde{Q}_{\ell_i,\alpha_{i}} \!+\! N_v ]_N}  x[q]   e^{j 2 \pi \left[ c_1 \ell_{i}^2 \!-\! q \ell_{i}/N \!+\! c_2 (q^2 \!-\! p^2) \right] } \\ \label{35}
	 & \times \mathcal{F}_i \left( p , q \right) + \omega [p].
\end{align}

From \eqref{35}, we observe that only partial $x[q]$ contribute to $y[p]$, forming a significant sparse structure.

\textbf{Remark 3: } \emph{In specific, if the AFDM system satisfies the narrowband approximation conditions mentioned in \textbf{Remark 1}, i.e.} $B \ll f_c $ \emph{and} $\alpha_i \ll 1/N$, \emph{we can obtain the following approximation for any $i$-th path,}
\begin{align}\label{36}
	Q_{\ell_i,\alpha_i} \approx \tilde{Q}_{\ell_i,\alpha_i} \approx p + 2 N c_1 \ell_{i}  - k_{i}
\end{align}
\emph{and}
\begin{align}\nonumber
	\mathcal{F}_i \left( p , q \right) & \approx \sum_{n=0}^{N-1} e^{-j 2 \pi \left(  p -  q  + 2 N c_1  \ell_{i} - k_{i} \right)  \frac{n}{N} } \\ \label{37}
	& = \frac{e^{-j 2 \pi \left( p - q + 2 N c_1  \ell_{i} - k_{i}\right) } - 1}{e^{-j \frac{2 \pi}{N} \left( p - q + 2 N c_1  \ell_{i} - k_{i}\right) } - 1}.
\end{align}


\emph{The input-output relationship in \eqref{35} can be reduced to}
\begin{align}\nonumber
	y \left[ p \right] \approx & \sum_{i=1}^{P} \frac{1}{N} {h}_i \sum_{q = [ Q_{\ell_i,\alpha_i} \!-\! N_v ]_N }^{[ Q_{\ell_i,\alpha_i} \!+\! N_v ]_N }  x[q]  e^{j 2 \pi \left[ c_1 \ell_{i}^2 \!-\! q \ell_{i}/N \!+\! c_2 (q^2 \!-\! p^2) \right] } \\ \label{38}
	& \times \frac{e^{-j 2 \pi \left( p - q + 2 N c_1  \ell_{i} - k_{i}\right) } - 1}{e^{-j \frac{2 \pi}{N} \left( p - q + 2 N c_1  \ell_{i} - k_{i}\right) } - 1} + \omega [p],
\end{align}
\emph{which is consistent with the input-output relationship of narrowband AFDM systems  mentioned in \cite{d9}.}


\section{AFDM Chirp Parameters Optimization}
From \eqref{33}, we observe that the wideband DAF domain equivalent channel representation depends on AFDM chirp parameters $c_1$ and $c_2$, where $c_1$ affects the DAF domain equivalent channel sparsity structure and $c_2$ only affects the phase shift. Therefore, we mainly focus on optimizing the chirp parameter $c_1$ to exploit the underlying wideband channel diversity, while $c_2$ is selected as an arbitrary irrational number as mentioned in \cite{d9}. 

According to \eqref{32}, we can obtain $Q_{\ell_i,\alpha_{max}} < Q_{\ell_i,-\alpha_{max}}$ and $\tilde{Q}_{\ell_i,\alpha_{max}} < \tilde{Q}_{\ell_i,-\alpha_{max}}$. 
For any $\ell_i \in [0, \ell_{max}]$, $\alpha_i \in \left[ - \alpha_{max} , \alpha_{max} \right]$ and  $k_i \in \left[ - k_{max} , k_{max} \right]$, the any $p$-th row non-zero elements range of $  \mathbf{H}_i $ can be given by
\begin{align}\label{39}
	q \in [ Q_{\ell_i,\alpha_{max}} - N_v , \tilde{Q}_{\ell_i , - \alpha_{max}} + N_v  ]_N .
\end{align}

Then, our target is to ensure that the non-zero entries in $\mathbf{H}_i$ and $\mathbf{H}_j$ do not overlap for any $i$ and $j$. In other words, the intersection of the non-zero entries between the $i$-th path and the  $j$-th path should be empty, i.e.,
\begin{align}\nonumber
	& [ Q_{\ell_i,\alpha_{max}} - N_v , \tilde{Q}_{\ell_i , - \alpha_{max}} + N_v ]_N  ~ \cap \\ \label{40}
	& [ Q_{\ell_j,\alpha_{max}} - N_v , \tilde{Q}_{\ell_j , - \alpha_{max}} + N_v ]_N  = \emptyset ,
\end{align}
where $i, j \in \left\lbrace 1, ... , P \right\rbrace$ and $i \neq j$. Here, we assume that $\ell_i < \ell_j$, then the constraint \eqref{40} can be converted to the following inequalities,
\begin{subequations}\label{41}
	\begin{align}
		Q_{\ell_j,\alpha_{max}} - N_v & > \tilde{Q}_{\ell_i , - \alpha_{max}} + N_v ,\label{41a} \\ 
		(\tilde{Q}_{\ell_{max} , - \alpha_{max}} + N_v)  & - (Q_{0,\alpha_{max}} - N_v) < N. \label{41b}
	\end{align}
\end{subequations}

Under the constraint \eqref{41a}, we can substitute \eqref{32} into \eqref{41a}, and the AFDM chirp parameter $c_1$ should satisfy
\begin{align}\label{42}
	c_1 > \frac{ 2 k_{{max}}  +  2 p \alpha_{max} +  2 (1 - \alpha_{max}^2) N_v }{  2 N[ \left( 1 \!-\! \alpha_{max}^2 \right)( \ell_{{j}} \!-\! \ell_{{i}}) \!-\! 2 \alpha_{max} \left( 2 \!-\! \alpha_{max}^2 \right) \left( N \!-\! 1 \right) ] }.
\end{align}

To ensure \eqref{42} always holds for any $p \in [0, N-1]$, $\ell_i$ and $\ell_j$ with $\ell_i < \ell_j$, while minimizing $c_1$ to achieve a more sparser DAF domain equivalent channel representation, the AFDM chirp parameter $c_1$ can be set as
\begin{align}\label{43}
	c_1 \!=\! \frac{2 k_{{max}} \!+\!  2 \alpha_{max} (N-1)  \!+\!  2 (1 \!-\! \alpha_{max}^2) N_v  \!+\! 1 }{ 2 N[ \left( 1 \!-\! \alpha_{max}^2 \right)( \ell_{\tilde{j}} \!-\! \ell_{\tilde{i}}) \!-\! 2 \alpha_{max} \left( 2 \!-\! \alpha_{max}^2 \right) \left( N \!-\! 1 \right) ] },
\end{align}
where
\begin{align}\label{44}
	\tilde{i},\tilde{j} = \arg\min\limits_{i,j} \left( \ell_j - \ell_i\right), \ell_i < \ell_j .
\end{align}

Specially, we assume that $\min(\ell_{{j}} - \ell_{{i}}) = 1$ for the scenarios with dense delays and ignore the term of $\alpha_{max}^2$, since the Doppler scale factor $\alpha_{max}$ is typically a small value in practice.\footref{n3} we can further simplify \eqref{43}  as
\begin{align}\label{45}
	c_1 \approx \frac{2 k_{{max}}  + 2 \alpha_{max} \left( N - 1 \right) + 2 N_v + 1 }{ 2N [1 - 4 \alpha_{max} \left( N - 1 \right)] }.
\end{align}

In addition, to ensure that $c_1 > 0$\footref{n2} always hold for \eqref{45}, the number of transmitted symbols $N$ should satisfy 
\begin{align}\label{46}
	N < \frac{1 }{ 4 \alpha_{max}} + 1.
\end{align}

On the other hand, under the constraint \eqref{41b}, we can substitute \eqref{32} into \eqref{41b}, and the AFDM chirp parameter $c_1$ should satisfy
\begin{align}\label{47}
	c_1 < \frac{ \left( 1 - \alpha_{max}^2 \right) (N - 2N_v) - 2 k_{{max}  }- 2\alpha_{max} \left( N - 1 \right) }{ 2N[\left( 1 -\alpha_{max}^2 \right) \ell_{{max}} + 2 \alpha_{max} \left( 2 - \alpha_{max}^2 \right)  \left( N - 1 \right)]}.
\end{align}

We again ignore the term $\alpha_{max}^2$ and further simplify \eqref{47} as
\begin{align}\label{48}
	c_1 < \frac{ N - 2 k_{{max}  }- 2\alpha_{max} \left( N - 1 \right) - 2 N_v}{2N [\ell_{{max}} + 4 \alpha_{max} \left( N - 1 \right)]}.
\end{align}

By substituting \eqref{45} into \eqref{48}, we can obtain the constraint relationship with the number of transmitted symbols $N$ as
\begin{align}\label{49}
	a_N N^2 + b_N N + c_N < 0,
\end{align}
where
\begin{subequations}\label{50}
	\begin{align}
		& a_N \!=\! 4 \alpha_{max}, \\ 
		& b_N \!=\! 2 \alpha_{max} \ell_{max} \!-\! 1,  \\ 
		& c_N \!=\! (2 k_{max} \!+\! 2 N_v) (\ell_{max} \!+\! 1) \!+\!  \ell_{max} \!-\! (2 \ell_{max} \!+\! 6) \alpha_{max}. 
	\end{align}
\end{subequations}

Note that \eqref{49} is a quadratic inequality about variable $N$. Then, the range of parameter $N$ with constraint \eqref{46} and \eqref{49} can be given by
\begin{align}\label{51}
	x_{L,N}  < N < \min \left( x_{H,N}, \frac{1}{4 \alpha_{max}} + 1\right), 
\end{align}
where
\begin{subequations}\label{52}
	\begin{align}
		x_{L,N} & = \frac{-b_N - \sqrt{b_N ^2 - 4 a_N c_N}}{2 a_N} , \\ 
		x_{H,N} & = \frac{-b_N + \sqrt{b_N ^2 - 4 a_N c_N}}{2 a_N} .
	\end{align}
\end{subequations}

The Doppler scale factor $\alpha_{max}$ is typically a small value in practice.\footref{n3} Consequently, the coefficients of the quadratic polynomial in \eqref{50} satisfy $a_N > 0, b_N < 0$ and $c_N > 0$. Therefore, $x_{L,N} > 0 $ always holds. Additionally, to ensure that \eqref{52} always has real solutions, we should satisfy the constraint
\begin{align}\label{53}
	b_N^2 - 4 a_N c_N > 0.
\end{align}

Note that \eqref{51} and \eqref{53} impose the constraints for the number of transmitted symbols $N$ and  channel spread characteristics, respectively. These are primarily due to the fact that the non-zero entries width $\mathcal{L}_i$ of the DAF domain equivalent channel in \eqref{34} is proportional to the number of transmitted symbols $N$ and Doppler scale factor $\alpha_{max}$, 
then it may fail to distinguish the channel paths 
for particularly large number of transmitted symbols  $N$ and large Doppler scale factor $\alpha_{max}$. 
Fortunately, the constraint \eqref{53} always holds in practice, and we can always determine a suitable number of  transmitted symbols $N$ for improving the AFDM system performance under time-scaled wideband doubly-dispersive channels. Without loss of generality, we take a typical time-scaled wideband underwater acoustic channel as an example, where the maximum Doppler factor $\alpha_{max} = 10^{-4}$ and maximum delay $\tau_{max} = 20$ms \cite{d39,d40,d41}. The carrier frequency is set as $f_c = 6$kHz with a transmission bandwidth $B = 4$kHz \cite{d41} and set $N_v = 2$. The constraint \eqref{53} can be calculated as $b_N ^2 - 4 a_N c_N \approx 0.17 > 0$. Therefore, the real solutions of $x_{L,N}$ and $x_{H,N}$ always exist in practice. Finally, to guarantee the underlying wideband channel diversity, we can set the AFDM chirp parameter $c_1$ as \eqref{45} and the number of transmitted symbols $N$ satisfies the constraint $\eqref{51}$.

\textbf{Remark 4:} \emph{In specific, if the AFDM system satisfies the narrowband approximation conditions mentioned in \textbf{Remark 1}, i.e.} $B \ll f_c $ \emph{and} $\alpha_i \ll  1/N$, \emph{the AFDM chirp parameter} $c_1$ \emph{in \eqref{45} can be simplified as}
\begin{align}\label{54}
	c_1 \approx \frac{2 k_{{max}} + 2 N_v + 1}{2N},
\end{align}
\emph{and the constraint \eqref{49} can be simplified as}
\begin{align}\label{55}
	\left(  2 k_{{max}} + 2 N_v \right)   \left( \ell_{{max}} + 1 \right)  + \ell_{{max}} < N,
\end{align}
\emph{which are consistent with the narrowband AFDM chirp parameter conditions} \emph{mentioned in \cite{d9}}.

\section{Performance Analysis}
Based on our optimized AFDM chirp parameter $c_1$ in \eqref{45} under time-scaled wideband doubly-dispersive channels, we analyze the correspondingly theoretical error performance based on PEP in this section. To facilitate the PEP analysis, \eqref{18} can be re-expressed as
\begin{align}\label{56}
	\mathbf{y} = \sum_{i=1}^{P} {h}_i \mathbf{H}_i \mathbf{x} + \boldsymbol{\omega} = \boldsymbol{\Phi} \left( \mathbf{x} \right) \mathbf{h} + \boldsymbol{\omega},
\end{align}
where 
\begin{subequations}\label{57}
	\begin{align}
		\mathbf{h} & = \left[ h_1, h_2, ..., h_P\right]^{\text{T}} \in \mathbb{C}^{P \times 1}, \\
		\boldsymbol{\Phi} \left( \mathbf{x} \right) & = \left[ \mathbf{H}_1 \mathbf{x}, \mathbf{H}_2 \mathbf{x}, ..., \mathbf{H}_P \mathbf{x}\right] \in \mathbb{C}^{N \times P}.
	\end{align}
\end{subequations}

The perfect channel state information (CSI) is assumed at the receiver. For a given time-scaled wideband doubly-dispersive channel, the conditional PEP of transmitted symbol $\mathbf{x}$ but erroneously decoded it as $\hat{\mathbf{x}}$ at receiver is given by
\begin{align}\label{58}
	P \left( \mathbf{x} \rightarrow \hat{\mathbf{x}} \left| \mathbf{h} \right.  \right) = Q \left( \sqrt{\frac{\left| \left[  \boldsymbol{\Phi} \left( \mathbf{x} \right) - \boldsymbol{\Phi} \left(  \hat{\mathbf{x}} \right) \right]   \mathbf{h} \right|^2  }{2 N_0}}\right), 
\end{align}
where $Q(x)$ represents the tail distribution function of a standard
Gaussian distribution. Here, we define $\boldsymbol{\Omega}_{ \mathbf{x}, \hat{\mathbf{x}}}  = \left[  \boldsymbol{\Phi} \left( \mathbf{x} \right) - \boldsymbol{\Phi} \left(  \hat{\mathbf{x}} \right) \right]^{\text{H}} \left[  \boldsymbol{\Phi} \left( \mathbf{x} \right) - \boldsymbol{\Phi} \left(  \hat{\mathbf{x}} \right) \right]  $, which is a Hermitian matrix and can be eigenvalue decomposed as $\boldsymbol{\Omega}_{ \mathbf{x}, \hat{\mathbf{x}}} = \mathbf{\tilde{U}}^{\text{H}} \boldsymbol{\tilde{\Lambda}} \mathbf{\tilde{U}} $, where $\mathbf{\tilde{U}} \in \mathbb{C}^{P \times P}$ is a unitary matrix and $\boldsymbol{\tilde{\Lambda}} = \text{diag} \left\lbrace \lambda_1, \lambda_2, ..., \lambda_P \right\rbrace $. 
The corresponding rank and the non-zero eigenvalues are defined as $R$ and $\lambda_i, i = 1, 2, ..., R$, respectively. We then have
\begin{align}\nonumber
	\left| \left[  \boldsymbol{\Phi} \left( \mathbf{x} \right) - \boldsymbol{\Phi} \left(  \hat{\mathbf{x}} \right) \right]   \mathbf{h} \right|^2  & = \mathbf{h}^{\text{H}} \boldsymbol{\Omega}_{ \mathbf{x}, \hat{\mathbf{x}}}  \mathbf{h} \\ \label{59}
	& =  \tilde{\mathbf{h}}^{\text{H}} \boldsymbol{\tilde{\Lambda}} \tilde{\mathbf{h}} = \sum_{i=1}^{R} \lambda_i \left| \tilde{h}_i \right|^2, 
\end{align}
where  $\tilde{\mathbf{h}} = \mathbf{\tilde{U}} \mathbf{h}$. According to $ Q \left( x \right) \leq \exp \left( 1/2 x^2 \right) , \forall x > 0 $, \eqref{58} can be further simplified as 
\begin{align}\label{60}
	P \left( \mathbf{x} \rightarrow \hat{\mathbf{x}} \left| \mathbf{h} \right.  \right) \leq \exp \left(- \frac{\sum_{i=1}^{R} \lambda_i \left| \tilde{h}_i \right|^2 }{4 N_0} \right). 
\end{align}

We assume that $h_i$ follows a Gaussian distribution with mean $0$ and variance $1/P$, i.e., $h_i \sim \mathcal{CN} \left( 0 , 1/P \right) $, such that $\tilde{h}_i$ also follows the distribution of $\mathcal{CN} \left( 0 , 1/P \right)$ since $\tilde{\mathbf{U}}$ is an unitary matrix. The final PEP can be calculated by averaging  channel statistics and given by
\begin{align}\label{61}
	P \left( \mathbf{x} \rightarrow \hat{\mathbf{x}}  \right) \leq \frac{1}{ \left(\frac{1}{4 N_0} \right)^R  \prod \limits_{i=1}^{R}  \frac{\lambda_i}{P}}.
\end{align}

From the above analysis, the average bit error rate (BER) can be upper bounded by
\begin{align}\label{62}
	P_e \leq \frac{1}{Q^{N} N \log_2Q} \sum_{\mathbf{x}}  \sum_{\mathbf{x} \neq \hat{\mathbf{x}}} P \left( \mathbf{x} \rightarrow \hat{\mathbf{x}}  \right) \mathbf{e} \left(\mathbf{x} , \hat{\mathbf{x}} \right), 
\end{align}
where $\mathbf{e} \left(\mathbf{x} , \hat{\mathbf{x}} \right)  $ represents the number of bits in difference for the corresponding pairwise error event. 

\begin{figure}[t!]
	\centering
	\includegraphics[scale=0.55]{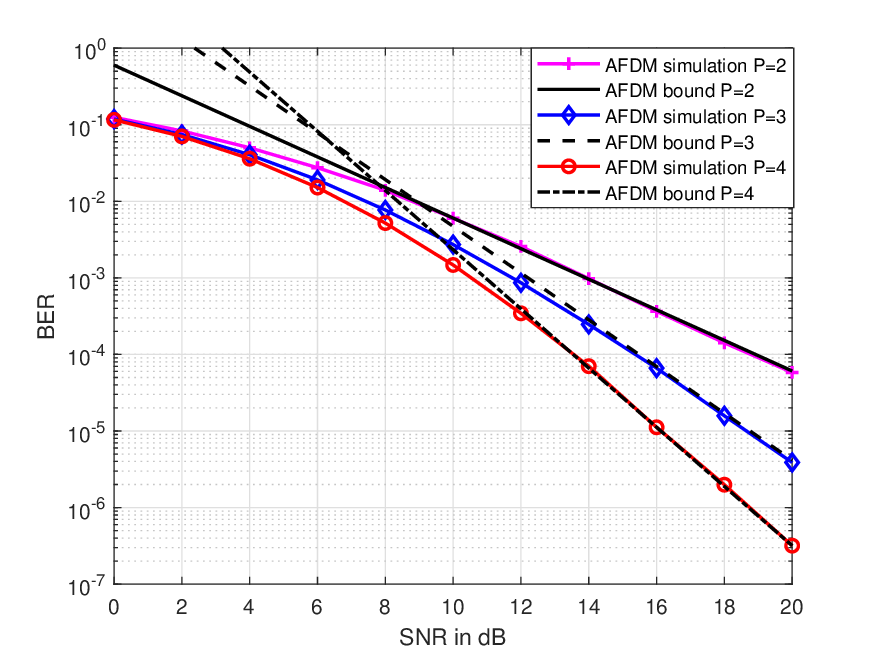}
	\caption{AFDM performance with different numbers of resolvable paths.}
	\label{fig_3}
\end{figure} 

Fig. \ref{fig_3} examines the AFDM performance based on ML detector with different numbers of resolvable paths in time-scaled wideband doubly-dispersive channels. Specifically, we consider a specific time-scaled wideband underwater acoustic channel with 
maximum Doppler scale factor $\alpha_{max} = 10^{-4}$ (after resampling processing \cite{d39,d40,d41})
and generate the complex exponential gains $h_i$ as independent Gaussian random variables with zero mean and variance $1/P$. The number of transmission symbols is $N = 16$ with BPSK modulation. AFDM chirp parameter $c_1$ is selected according to \eqref{45}. 
From Fig. \ref{fig_3}, we can observe that the AFDM performance bound and ML detection results tend to converge at high signal-to-noise ratio (SNR) for different numbers of resolvable paths, demonstrating the effectiveness of our PEP theoretical analysis. Additionally, the BER performance improves as the number of resolvable paths $P$ increases. This conclusion illustrates that a higher diversity gain can be achieved with an increasing number of resolvable paths $P$.
\begin{figure}[t!]
	\centering
	\includegraphics[scale=0.55]{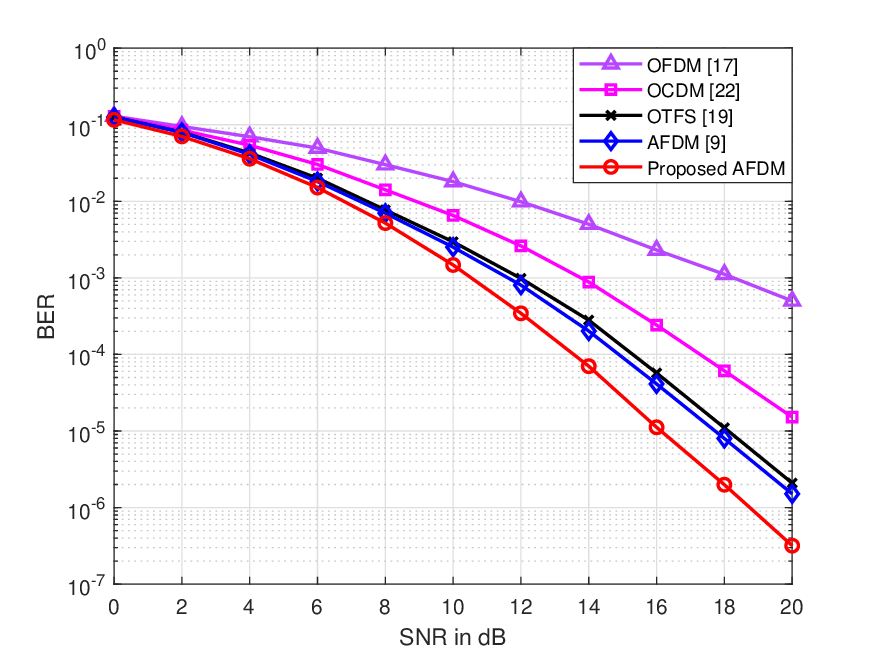}
	\caption{BER performance comparison for different modulation schemes.}
	\label{fig_4}
\end{figure}

We further test the BER performance with different modulation schemes based on ML detection in Fig. \ref{fig_4} under time-scaled wideband doubly-dispersive channels.\footnote{\label{n4}Note that all modulation schemes have been added the CPP and CPS for a fair comparison.} Specially, the delay dimension $\tilde{M}$ and Doppler dimension $\tilde{N}$ are $\tilde{M} = 4$ and $\tilde{N} = 4$ in OTFS systems.
As shown in Fig. 4, OFDM \cite{d16} cannot effectively resolve the channel paths, leading to worst BER performance. OCDM \cite{d21} can resolve the partial channel paths and exhibit better BER performance than OFDM. 
However, the different paths may still overlap under certain channel spread for OCDM, where the BER performance is highly dependent on the channel spread characteristics.
OTFS \cite{d18} and AFDM modulation with narrowband chirp parameters \cite{d9} can theoretically resolve all paths under narrowband doubly-dispersive channels, but it fails to distinguish paths in time-scaled wideband doubly-dispersive channels due to neglecting the time-scale variation caused by wideband Doppler effect. In addition, the AFDM with narrowband chirp parameters also shows a slight performance improvement compared to the OTFS modulation. This is mainly due to the fact that the transmitted AFDM symbols experience less interference from time-scaling effects compared to OTFS, which benefit from the underlying one dimensional transform structure compared to the two dimensional transform structure of OTFS modulation.
Finally, our optimized chirp parameters in AFDM system can effectively resolve the different paths and exploit the underlying wideband channel diversity for performance enhancement. Such results verify that our optimized chirp parameters are effective in time-scaled wideband AFDM transmissions. 

\section{Receiver Design}
\begin{figure}[t!]
	\captionsetup{font = {scriptsize,scriptsize,scriptsize,small}, format = hang, justification = centering}
	\subfloat[DAF domain $|\bar{\mathbf{H}}|$.]{\includegraphics[scale=0.3]{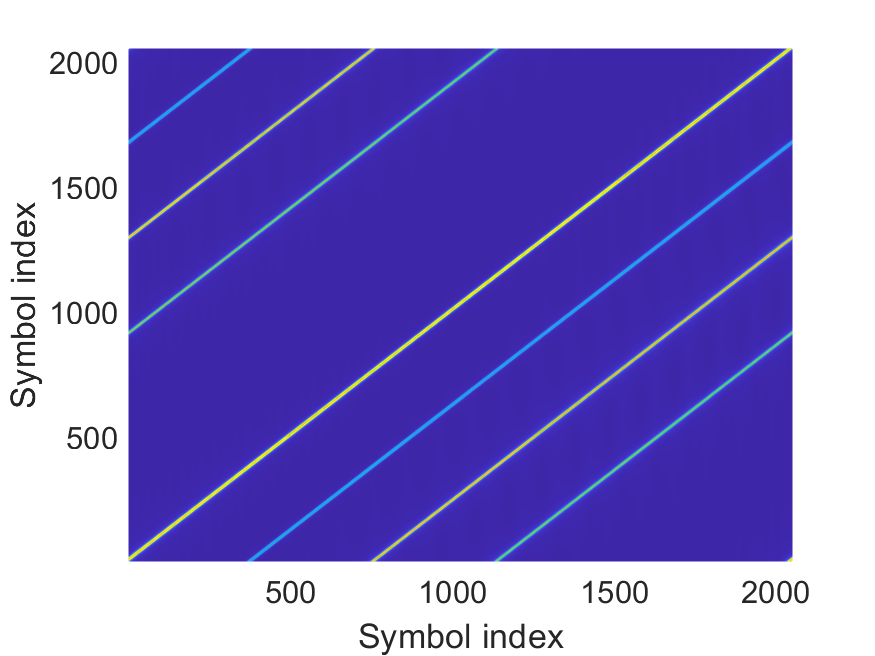}}
	\subfloat[Time domain $|\bar{\mathbf{H}}_{\text{T}}|$.]{\includegraphics[scale=0.3]{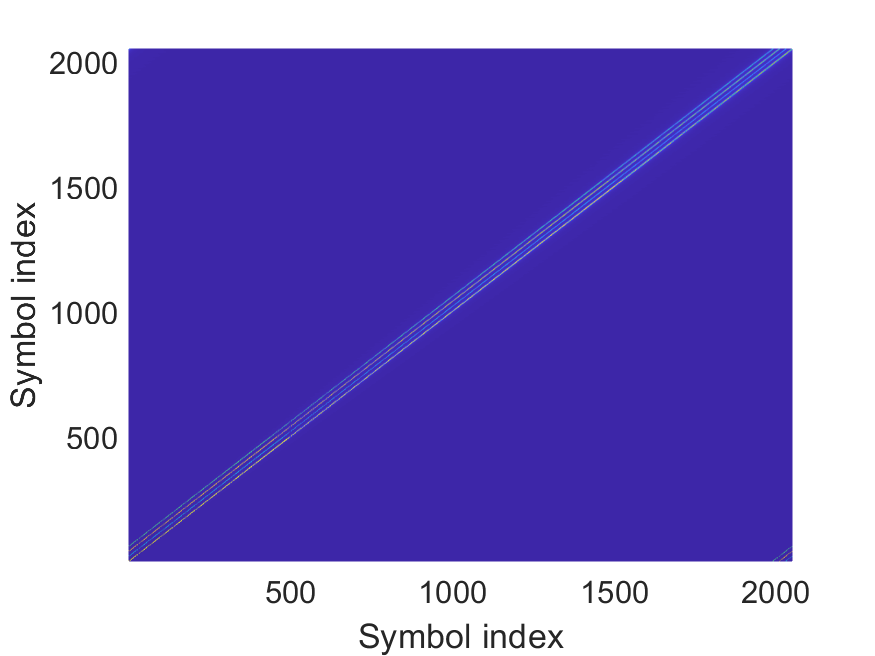}}
	\hfil
	\caption{Equivalent channel structure for DAF domain and time domain.}
	\label{fig_5}
\end{figure}

Although the ML detector can achieve the optimal BER performance, it is impractical due to its exponentially growing complexity with the system dimensions. From \eqref{33}, we observe that the DAF domain equivalent channel exhibits a significant sparse structure, which facilitates the application of reduced complexity sparse detection algorithms. However, 
the sparse DAF domain channel will be very dense with increasing the number of transmission symbols $N$ and Doppler scale factor $\alpha$ according to \eqref{34}, leading to a high detection complexity even with sparse detection algorithms. To tackle this issue, we focus on the sparser time domain channel for reduced complexity receiver design. 
To intuitively illustrate the sparse characteristics of the time domain and DAF domain equivalent channel, we present the corresponding equivalent channel structure in Fig. \ref{fig_5}.
It is obvious that the time domain equivalent channel $\bar{\mathbf{H}}_{\text{T}}$ has a more sparser structure than DAF domain equivalent channel $\bar{\mathbf{H}}$ under time-scaled wideband doubly-dispersive channels. 
This observation motivates us to consider the more sparser time domain equivalent channel for reduced complexity detector design.


\subsection{Cross Domain Distributed OAMP detector}
According to \eqref{14} and \eqref{18}, we can decouple the problem of AFDM symbol detection into the following constraints, i.e.,
\begin{subequations}\label{63}
	\begin{align}
		\text{Linear constraint}  ~\{\boldsymbol{\Gamma}\} &: \mathbf{y}_{\text{T}} = \bar{\mathbf{H}}_{\text{T}} \mathbf{x}_{\text{T}} + \boldsymbol{\omega}_{\text{T}}, \label{63a} \\ 
		\text{Non-linear constraint} ~ \{\boldsymbol{\Phi}\} &:  \mathbf{x} \in \mathbb{A}^{N \times 1}. \label{63b}
	\end{align}
\end{subequations}

\begin{figure*}[t!]
	\centering
	\includegraphics[scale=0.68]{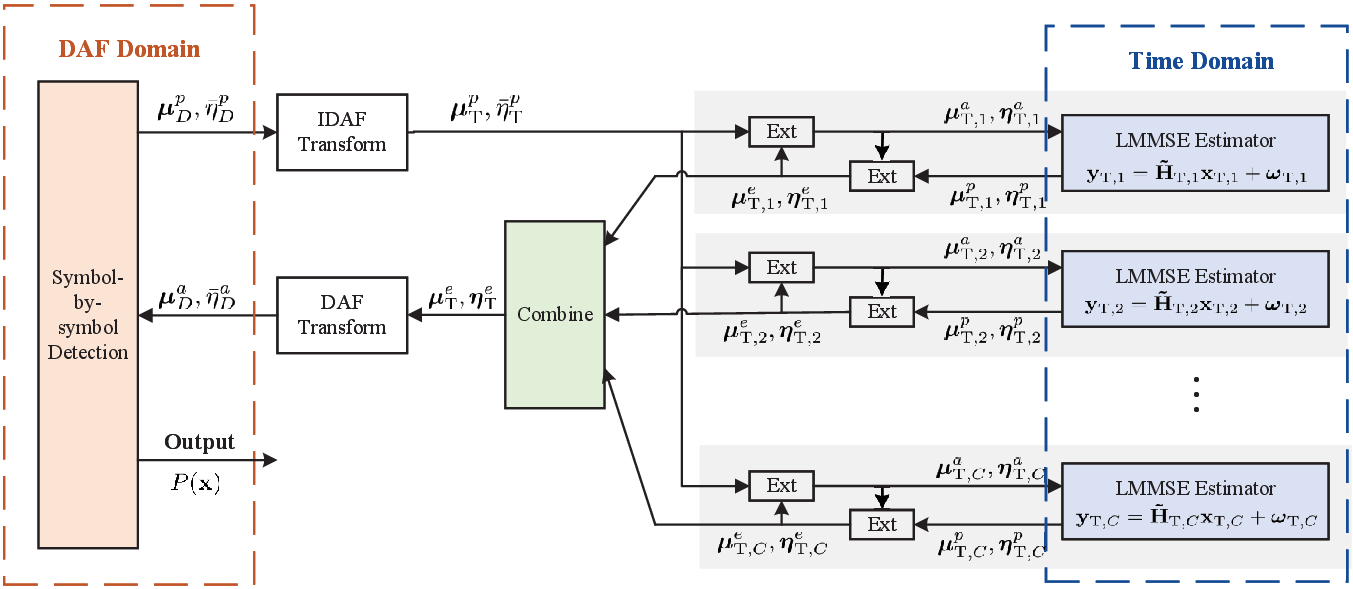}
	\caption{Proposed cross domain distributed detection structure.}
	\label{fig_6}
\end{figure*}

Then, we aim to find the MMSE estimation of $\mathbf{x}$. For this issue, we propose an efficient CD-D-OAMP detector, which is described in Fig. \ref{fig_6}. It iteratively optimize the linear module in time domain and the nonlinear module in DAF domain to converge towards overall MMSE estimation.
In addition, it can effectively leverage the sparse time domain channel structure for complexity reduction and achieve desired complexity and performance trade-off. The proposed CD-D-OAMP detection algorithm is summarized in \textbf{Algorithm 1} and the specific details are introduced as following.

\textbf{$\left. \boldsymbol{1} \right) $ Linear constraint} $\{\boldsymbol{\Gamma}\}$ \textbf{in time domain with distributed LMMSE (D-LMMSE) estimation}: Under the linear constraint in \eqref{63a}, we propose the D-LMMSE time domain estimator to leverage the sparse time domain channel structure for complexity reduction.
Specifically, we divide the time domain received signal $\mathbf{y}_{\text{T}}$ into $C$ groups and the number of each group received signal is $N_c = N / C$. When $ C = 1$, the D-LMMSE estimator reduces to the original LMMSE estimator. Then, the time domain received signal $\mathbf{y}_{\text{T}}$, channel matrix $\bar{\mathbf{H}}_{\text{T}}$ and noise $\boldsymbol{\omega}_{\text{T}}$ can be partitioned as
\begin{subequations}\label{65}
\begin{align}
	\mathbf{y}_{\text{T}} & = [\mathbf{y}_{\text{T},1}^{\text{T}}, \mathbf{y}_{\text{T},2}^{\text{T}}, ..., \mathbf{y}_{\text{T},C}^{\text{T}}]^{\text{T}}, \\
	\bar{\mathbf{H}}_{\text{T}} & = [\bar{\mathbf{H}}_{\text{T},1}^{\text{T}}, \bar{\mathbf{H}}_{\text{T},2}^{\text{T}}, ..., \bar{\mathbf{H}}_{\text{T},C}^{\text{T}}]^{\text{T}}, \\
	\boldsymbol{\omega}_{\text{T}} & = [\boldsymbol{\omega}_{\text{T},1}^{\text{T}}, \boldsymbol{\omega}_{\text{T},2}^{\text{T}}, ..., \boldsymbol{\omega}_{\text{T},C}^{\text{T}}]^{\text{T}},
\end{align}
\end{subequations}
where $ \mathbf{y}_{\text{T},c} \in \mathbb{C}^{N_c \times 1}$, $ \bar{\mathbf{H}}_{\text{T},c} \in \mathbb{C}^{N_c \times N}$ and $ \boldsymbol{\omega}_{\text{T},c} \in \mathbb{C}^{N_c \times 1}$. The $c$-th group received signal $\mathbf{y}_{\text{T},c}$ can be given by
\begin{align}\label{66}
	\mathbf{y}_{\text{T},c} = \bar{\mathbf{H}}_{\text{T},c} \mathbf{x}_{\text{T}} + \boldsymbol{\omega}_{\text{T},c} , c = 1, 2, ..., C.
\end{align}

\begin{figure}[t!]
	\centering
	\includegraphics[scale=0.4]{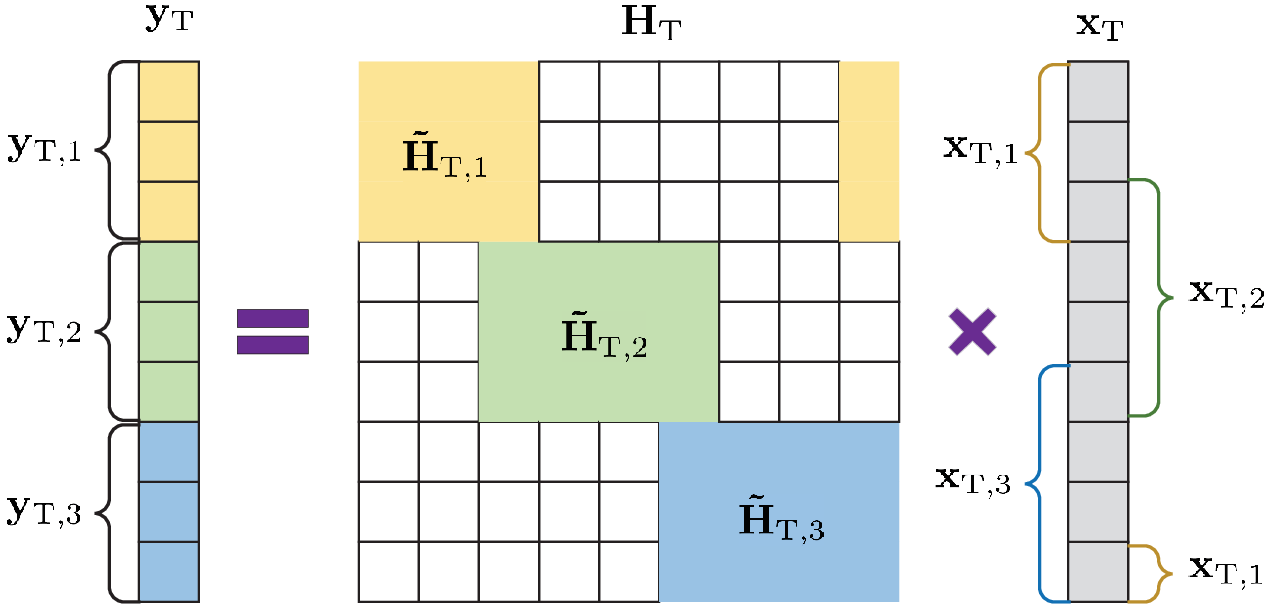}
	\caption{Distributed input-output relationship with three groups as an example.}
	\label{fig_7}
\end{figure}

Due to the sparse structure of the time domain equivalent channel matrix, only a small subset $\mathbf{x}_{\text{T},c} = \mathbf{x}_{\text{T}} [\mathcal{D}_c] \in \mathbb{C}^{|\mathcal{D}_c| \times 1} $ is associated with $\mathbf{y}_{\text{T},c}$, where $\mathcal{D}_c$ is the index set of $\mathbf{x}_{\text{T}}$ correlated to $\mathbf{y}_{\text{T},c}$ and $|\mathcal{D}_c| $ is the size of $\mathcal{D}_c$. Therefore, we can only retain the column elements with indices $\mathcal{D}_c$ from $\bar{\mathbf{H}}_{\text{T},c}$ and form a  smaller-dimension channel matrix $\mathbf{\tilde{H}}_{\text{T},c} \in \mathbb{C}^{N_c \times |\mathcal{D}_c|}$, i.e.,  $\mathbf{\tilde{H}}_{\text{T},c} = \bar{\mathbf{H}}_{\text{T},c} [:, \mathcal{D}_c]$. Typically, $|\mathcal{D}_c|$ is much smaller than $N$ due to the sparse property of the time domain equivalent channel. Fig. \ref{fig_7} intuitively illustrates the distributed time domain input-output relationship with three groups as an example. We can further simplify \eqref{66} as 
\begin{align}\label{67}
	\mathbf{y}_{\text{T},c} = \mathbf{\tilde{H}}_{\text{T},c} \mathbf{x}_{\text{T},c} + \boldsymbol{\omega}_{\text{T},c}.
\end{align}


According to LMMSE criterion, the posteriori mean $\boldsymbol{\mu}_{\text{T},c}^{p,(\iota)}$ and variance $\boldsymbol{\Sigma}_{\text{T},c}^{p,(\iota)}$ of $\mathbf{x}_{\text{T},c}$ in $(\iota)$-th iteration are given by
\begin{subequations}\label{68}
	\begin{align}
		\boldsymbol{\mu}_{\text{T},c}^{p,(\iota)} & = \boldsymbol{\mu}_{\text{T},c}^{a,(\iota)} + \mathbf{W}^{(\iota)}_{\text{T},c} \left( \mathbf{y}_{\text{T},c} - \mathbf{\tilde{H}}_{\text{T},c}  \boldsymbol{\mu}_{\text{T},c}^{a,(\iota)} \right), \label{68a} \\ 
		\boldsymbol{\eta}_{\text{T},c}^{p,(\iota)} & = \text{diag} \left\lbrace \boldsymbol{\Sigma}^{a,(\iota)}_{\text{T},c} - \mathbf{W}^{(\iota)}_{\text{T},c} \mathbf{\tilde{H}}_{\text{T},c} \boldsymbol{\Sigma}^{a,(\iota)}_{\text{T},c} \right\rbrace, \label{68b}
	\end{align}
\end{subequations}
where $\mathbf{W}^{(\iota)}_{\text{T},c} $ is the $c$-th group time domain LMMSE filter,
\begin{align}\label{69}
	\mathbf{W}^{(\iota)}_{\text{T},c} = \boldsymbol{\Sigma}^{a,(\iota)}_{\text{T},c} \mathbf{\tilde{H}}^{\text{H}}_{\text{T},c} \left( \mathbf{\tilde{H}}_{\text{T},c} \boldsymbol{\Sigma}^{a,(\iota)}_{\text{T},c} \mathbf{\tilde{H}}^{\text{H}}_{\text{T},c} + \mathbf{I}_{N_c} N_0 \right)^{-1},
\end{align}
and $ \boldsymbol{\Sigma}^{a,(\iota)}_{\text{T},c} = \text{diag} \left\lbrace \boldsymbol{\eta}^{a,(\iota)}_{\text{T},c} \right\rbrace $.
The $\boldsymbol{\mu}_{\text{T},c}^{a,(\iota)}$ and $\boldsymbol{\eta}_{\text{T},c}^{a,(\iota)}$ are the priori mean and variance in $(\iota)$-th iteration, which can be initialized as  $\boldsymbol{\mu}_{\text{T},c}^{a,(1)} = \boldsymbol{0}_{N_c} $ and $\boldsymbol{\eta}_{\text{T},c}^{a,(1)} = \boldsymbol{1}_{N_c} $ in the first iteration.

Then, the extrinsic mean $\boldsymbol{\mu}_{\text{T},c}^{e,(\iota)}$ and variance $\boldsymbol{\eta}_{\text{T},c}^{e,(\iota)} $ can be obtained by
\begin{subequations}\label{70}
	\begin{align}
		{\eta}_{\text{T},c}^{e,(\iota)} [d]   & = \left( \frac{1}{{\eta}_{\text{T},c}^{p,(\iota)} [d]}  -  \frac{1}{{\eta}_{\text{T},c}^{a,(\iota)} [d]} \right)^{-1}  , \label{70a} \\
		{\mu}_{\text{T},c}^{e,(\iota)} [d] & = {\eta}_{\text{T},c}^{e,(\iota)} [d]  \left( \frac{{\mu}_{\text{T},c}^{p,(\iota)} [d]}{{\eta}_{\text{T},c}^{p,(\iota)} [d]}  -  \frac{{\mu}_{\text{T},c}^{a,(\iota)} [d]}{{\eta}_{\text{T},c}^{a,(\iota)} [d]} \right), \label{70b}
	\end{align}
\end{subequations}
where $d = 1, 2, ..., |\mathcal{D}_c|$.
Following the Gaussian message combining rule, we can update the extrinsic marginal distribution $P_{E}^{(\iota)} \left( \mathbf{x}_{\text{T}} \right)  \sim \mathcal{CN} \left( \boldsymbol{\mu}_{\text{T}}^{e,(\iota)} , \boldsymbol{\eta}_{\text{T}}^{e,(\iota)} \right) $ with
\begin{subequations}\label{71}
	\begin{align}
		{\eta}_{\text{T}}^{e,(\iota)} [j]  & = \left( \sum_{\bar{c}_j \in \bar{C}_j}  \frac{1}{{\eta}_{\text{T},\bar{c}_j}^{e,(\iota)} [\mathcal{I}_{j}]}   \right)^{-1}, \label{71a} \\
		{\mu}_{\text{T}}^{e,(\iota)} [j]  & = {\eta}_{\text{T}}^{e,(\iota)} [j] \left( \sum_{\bar{c}_j \in \bar{C}_j} \frac{{\mu}_{\text{T},\bar{c}_j}^{e,(\iota)} [\mathcal{I}_{j}]}{{\eta}_{\text{T},\bar{c}_j}^{e,(\iota)} [\mathcal{I}_{j}]} \right), \label{71b}
	\end{align}
\end{subequations}
where $j = 1, 2, ..., N$, $\bar{c}_j \in \bar{C}_j$ is the index of $j$-th detection symbol associated with $\mathbf{x}_{\text{T},\bar{c}_j}$ and $\mathcal{I}_{j}$ is the corresponding position in $\bar{c}_j$-th group, $\bar{C}_j$ is the index set of $\bar{c}_j$.


Then, we can transform the time domain extrinsic mean $\boldsymbol{\mu}_{\text{T}}^{e,(\iota)} $ and variance $\boldsymbol{\eta}_{\text{T}}^{e,(\iota)} $ into DAF domain as the \emph{a priori}  mean $\boldsymbol{\mu}_{D}^{a,(\iota)}$ and variance $\boldsymbol{\eta}_{D}^{a,(\iota)}$ by DAF transform $\mathbf{U}$, i.e.,
\begin{align}\label{72}
	\boldsymbol{\mu}_{D}^{a,(\iota)} = \mathbf{U} \boldsymbol{\mu}_{\text{T}}^{e,(\iota)},  
	\boldsymbol{\eta}_{D}^{a,(\iota)} = \boldsymbol{\eta}_{T}^{e,(\iota)}.
\end{align}

Furthermore, we do the sample average of variance $\boldsymbol{\eta}_{D}^{a,(\iota)}$ as
\begin{align}\label{73}
	\bar{\eta}_{D}^{a,(\iota)}  = \frac{1}{N} \sum_{j=1}^{N} {\eta}_{\text{D}}^{a,(\iota)} [j]. 
\end{align}

Next, the priori mean $\boldsymbol{\mu}_{D}^{a,(\iota)}$ and average variance $\bar{\eta}_{D}^{a,(\iota)}$ are passed to the non-linear detection module.

\textbf{$\left. \boldsymbol{2} \right) $ Non-linear constraint $\{\boldsymbol{\Phi}\}$ in DAF domain with symbol-by-symbol detection}: The posteriori distribution of $j$-th DAF domain symbol can be given by
\begin{align}\label{74}
	\bar{P}^{(\iota)} \left( x_j = \chi \right) \propto P_D \left( x_j = \chi \right) \exp \left(- \frac{\left| \chi - {\mu}_{D}^{a,(\iota) } [j] \right|^2 }{\bar{\eta}_{D}^{a,(\iota)}}  \right),
\end{align}
where $\chi \in \mathbb{A}$. $ P_D \left( x_j = \chi \right)$ denotes the \emph{a priori} probability and is also assumed to be equiprobable symbols if no priori information observed. Then, we can project the $j$-th symbol posteriori probability $\bar{P}^{(\iota)} (x_j) $ into a Gaussian distribution $\mathcal{CN} \left( {\mu}_{D}^{p,(\iota) } [j], {\eta}_{D}^{p,(\iota) } [j] \right) $ with
\begin{subequations}\label{75}
	\begin{align}
		{\mu}_{D}^{p,(\iota) } [j] & = \sum_{\chi \in \mathbb{A}} \chi \bar{P}^{(\iota)} \left( x_j = \chi \right), \\
		{\eta}_{D}^{p,(\iota) } [j] & = \sum_{\chi \in \mathbb{A}} \bar{P}^{(\iota)} \left( x_j = \chi \right) \left| \chi - {\mu}_{D}^{p,(\iota) } [j] \right|^2. 
	\end{align}
\end{subequations}

Similarly, we further do the sample average of the variance $\boldsymbol{\eta}_{D}^{p,(\iota) }$, i.e., 
\begin{align}\label{76}
	\bar{\eta}_{D}^{p,(\iota) }  = \frac{1}{N}\sum_{{j}=1}^{N} {\eta}_{D}^{p,(\iota) } [{j}].
\end{align}


The DAF domain posteriori mean $\boldsymbol{\mu}_{\text{D}}^{p,(\iota)} $ and average variance $\bar{\eta}_{\text{D}}^{p,(\iota)} $ can be transformed into time domain by IDAF transform $\mathbf{U}^{\text{H}}$, i.e.,
\begin{align}\label{77}
	\boldsymbol{\mu}_{\text{T}}^{p,(\iota)} = \mathbf{U}^{\text{H}} \boldsymbol{\mu}_{\text{D}}^{p,(\iota)}, 
	\bar{\eta}_{\text{T}}^{p,(\iota)} = \bar{\eta}_{\text{D}}^{p,(\iota)}.
\end{align}

The priori mean $\boldsymbol{\mu}_{\text{T},c}^{a,(\iota+1)}$ and variance $\boldsymbol{\eta}_{\text{T},c}^{a,(\iota+1)}$ in $c$-th group at $(\iota+1)$-th iteration can be obtained by
\begin{subequations}\label{78}
	\begin{align}
		{\eta}_{\text{T},c}^{a,(\iota+1)} [d] & = \left( \frac{1}{\bar{\eta}_{\text{T}}^{p,(\iota)}   }  - \frac{1}{\eta_{\text{T},c}^{e,(\iota)}[d] }  \right)^{-1}, \label{78a} \\
		{\mu}_{\text{T},c}^{a,(\iota+1)} [d] & \!=\! {\eta}_{\text{T},c}^{a,(\iota+1)} [d]  \left(  \frac{{\mu}_{\text{T}}^{p,(\iota)} [\mathcal{D}_c (d)] }{\bar{\eta}_{\text{T}}^{p,(\iota)}} \!-\!  \frac{{\mu}_{\text{T},c}^{e,(\iota)} [d] }{\eta_{\text{T},c}^{e,(\iota)} [d]} \right). \label{78b}
	\end{align}
\end{subequations}

Finally, we can pass the priori mean $\boldsymbol{\mu}_{\text{T},c}^{a,(\iota+1)}$ and variance $\boldsymbol{\eta}_{\text{T},c}^{a,(\iota+1)}$ of $c$-th group to the  linear module and form the iterative loop.

\textbf{$\left. \boldsymbol{5} \right) $ Convergence indicator}: When all DAF domain symbols are updated, we can compute the convergence indicator $\theta^{\left( \iota \right) } $ by
\begin{equation}\label{79}
	\theta^{\left(\iota \right) } = \frac{1}{N} \sum_{j = 1}^{N} \mathbb{I} \left(  \mathop{\max}\limits_{\chi \in \mathbb{A}} {\bar{P}^{\left( \iota \right) } \left( x_j = \chi \right)} \geq 1 - \varrho \right)
\end{equation}
with a small $\varrho > 0$ and $\mathbb{I} \left( \cdot \right) $ is the indicator function.

\textbf{$\left. \boldsymbol{6} \right) $ Update criterion}: If $\theta^{\left( \iota \right)} \geq \theta^{\left( \iota-1 \right) }$, we can update
\begin{equation}\label{80}
	{{P}} \left( x_j \right)  = \bar{{P}}^{\left( \iota \right) } \left( x_j \right), j = 1, 2, ..., N.
\end{equation}

\textbf{$\left. \boldsymbol{7} \right) $ Stopping criterion:} The CD-D-OAMP detector terminates when either $\theta^{\left( \iota \right) } = 1$ or the maximum iteration number $n_{t}$ is reached. Then, we can make the final decisions as
\begin{equation}\label{81}
	\hat{x}_j = \mathop{\arg \max}\limits_{\chi \in \mathbb{A}} {P} \left( x_j = \chi \right), j = 1, 2, ..., N.
\end{equation}

\textbf{Remark 5}: \emph{Note that both the D-LMMSE estimation in time domain and the symbol-by-symbol detection in DAF domain can support parallel computing processes, which can significantly reduce the computational latency. When C=1, our proposed CD-D-OAMP detector can be reduced to cross domain OAMP (CD-OAMP) detector for AFDM system. In addition, our proposed CD-D-OAMP detector can also be simplified as the distributed OAMP (D-OAMP) detector by directly performing D-LMMSE estimation and symbol-by-symbol detection in DAF domain. Furthermore, the D-OAMP detector is reduced to the OAMP detector when $C=1$ \cite{d31}.}


\begin{algorithm}[t!]
	\caption{CD-D-OAMP Iterative Detector }\label{alg1}
	Input: $\mathbf{y}_{\text{T}}$, $\bar{\mathbf{H}}_{\text{T}}$,  $n_{t}$, $ {P}_{D} \left( \mathbf{x} \right) $\\
	Initialize: $\boldsymbol{\mu}_{\text{T},c}^{a,(1)} = \boldsymbol{0}_{N_c} $, $\boldsymbol{\eta}_{\text{T},c}^{a,(1)} = \boldsymbol{1}_{N_c} $ and $\iota = 1$, $c = 1, 2, ..., C$ \\
	\textbf{Repeat}
	\begin{algorithmic}[1]
		\STATE Perform the D-LMMSE estimation and generate the time domain posteriori mean $\boldsymbol{\mu}_{\text{T},c}^{p,(\iota)}$ and variance $\boldsymbol{\eta}_{\text{T},c}^{p,(\iota)}$ by \eqref{68}, $c = 1, 2, ..., C$;
		\STATE Generate the time domain extrinsic mean $\boldsymbol{\mu}_{\text{T}}^{e,(\iota)}$ and variance $\boldsymbol{\eta}_{\text{T}}^{e,(\iota)}$ by \eqref{71};
		\STATE Transform the time domain extrinsic mean and variance into DAF domain as priori mean $\boldsymbol{\mu}_{D}^{a,(\iota)}$ and variance $\bar{\eta}_{D}^{a,(\iota)}$ by \eqref{72} and \eqref{73};
		\STATE Perform the symbol-by-symbol detection in DAF domain and generate posteriori mean $\boldsymbol{\mu}_{D}^{p,(\iota)}$ and variance $\bar{\eta}_{D}^{p,(\iota)}$ by \eqref{75} and \eqref{76};
		\STATE Transform the DAF domain posteriori mean and variance into the time domain mean $\boldsymbol{\mu}_{\text{T}}^{p,(\iota)}$ and variance $\bar{\eta}_{\text{T}}^{p,(\iota)}$ in \eqref{77};
		\STATE Generate the $c$-th group time domain priori mean $\boldsymbol{\mu}_{\text{T},c}^{a,(\iota+1)}$ and variance $\boldsymbol{\eta}_{\text{T},c}^{a,(\iota+1)}$  by \eqref{78}, $c = 1, 2, ..., C$;
		\STATE Compute the convergence indicator in \eqref{79};
		\STATE Update $P (\mathbf{x}) = \bar{P}^{(\iota)} (\mathbf{x})$ in \eqref{80};
		\STATE  $ \iota = \iota + 1 $;
	\end{algorithmic}
	\textbf{Until} : $\theta^{\left( \iota \right) } = 1$ or $\iota = n_{t}$; \\
	\textbf{Output}  : The decisions of the transmitted symbols in \eqref{81}.
	\label{alg1}
\end{algorithm} 

\subsection{MSE Performance Analysis via State Evolution}
In this sub-section, we investigate the average MSE performance of the proposed CD-D-OAMP detector based on state evolution. Since the time domain D-LMMSE estimator involves multi-group symbols estimation and combination, we track the time domain output variance of each group to better characterize the state evolution process. The finally state evolution can be characterized by
\begin{subequations}\label{82}
	\begin{align}
		&\!\bar{\eta}_{\text{D}}^{p,(\iota)}  \!=\! f_{\text{D}} \!\left(\! f_{\text{C}} \!\left(\! \left[\!  \text{diag} \!\left\lbrace\! f_{\text{T}} \!\left(\! \boldsymbol{\eta}_{\text{T},c}^{a,(\iota)} \!\right) \! \right\rbrace^{\!-1} \!-\! \text{diag} \!\left\lbrace \! \boldsymbol{\eta}_{\text{T},c}^{a,(\iota)} \! \right\rbrace^{\!-1} \! \right]^{\!-1} \!\right) \!\right)\!,\!   \\
		&\boldsymbol{\eta}_{\text{T},c}^{a,(\iota+1)}  \!=\! \left(\!  \frac{\boldsymbol{1}_{N_c}}{\bar{\eta}_{\text{D}}^{p,(\iota)}} \!-\! \left[ \! \text{diag} \! \left\lbrace \! f_{\text{T}} \! \left( \! \boldsymbol{\eta}_{\text{T},c}^{a,(\iota)} \! \right) \! \right\rbrace^{\!-\!1}  \!-\! \text{diag} \! \left\lbrace \! \boldsymbol{\eta}_{\text{T},c}^{a,(\iota)} \! \right\rbrace^{\!-\!1} \!\right]  \!\right)^{\!-\!1}
	\end{align}
\end{subequations}
with initialization $\boldsymbol{\eta}_{\text{T},c}^{a,(1)} = \boldsymbol{1}_{N_c} $, where $c = 1, 2, ..., C$. $f_{\text{C}} \left( \cdot \right) $ denotes the combination operation of D-LMMSE detection and is associated with trimmed channel matrix $\mathbf{\tilde{H}}_{\text{T},c}$, which can be given by \eqref{71a} and average by \eqref{73}. 
$f_{\text{T}} \left( \cdot \right) $ denotes the posteriori variance of time domain LMMSE estimator given the corresponding priori information, which can be obtained by \eqref{68b}. $f_{\text{D}} \left( \cdot \right) $ denotes the posteriori variance of symbol-by-symbol detection in DAF domain given the corresponding priori information, it can be regarded as the AWGN observation with corresponding priori variance. For example, for a given AWGN observation $\hat{x}$ with average noise variance $\bar{\eta}_{\text{D}}^{a,(\iota)} $, i.e.,
\begin{align}\label{83}
	\hat{x} = x + \sqrt{\bar{\eta}_{\text{D}}^{a,(\iota)}} z,
\end{align}
where $z$ is a Gaussian variable with distribution $\mathcal{CN} \left(0, 1\right) $. The posteriori variance $f_{\text{D}} \left( \bar{\eta}_{\text{D}}^{a,(\iota)} \right) $ of $x$ given $\hat{x}$ can be given by
\begin{align}\label{84}
	f_{\text{D}} \left( \bar{\eta}_{\text{D}}^{a,(\iota)} \right) = \mathbb{E} \left\lbrace \left| x - \mathbb{E} \left[ x | \hat{x}\right] \right|^2  \right\rbrace.
\end{align}

Note that \eqref{84} is  a nonlinear function with respect to the specific constellation shape $\mathbb{A}$ and it is challenge to obtain a theoretical expression. Thus, we consider using a Monte Carlo approach to approximate the average posteriori variance and verify the convergence of the proposed algorithm \cite{d38}.

\begin{figure}[t!]
	\centering
	\includegraphics[scale=0.55]{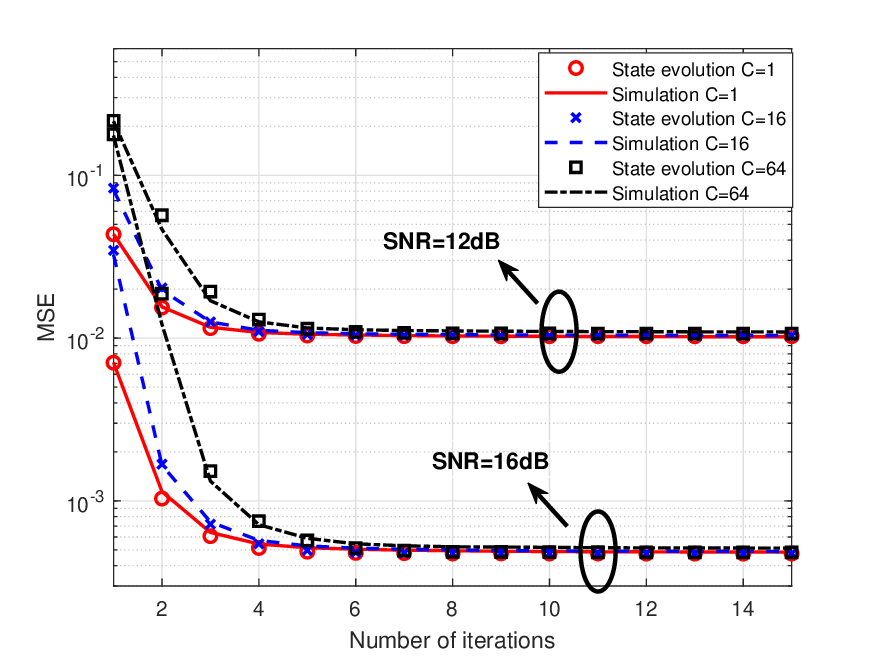}
	\caption{State evolution verification of the proposed CD-D-OAMP detector.}
	\label{fig_8}
\end{figure} 

From \textbf{Algorithm 1}, we observe that the output estimation of $\mathbf{\hat{x}}$ is produced from the symbol-by-symbol detection in DAF domain. Therefore, we track evolution process of the proposed CD-D-OAMP detector via average posteriori variance $\bar{\eta}_{D}^{p,(\iota)}$ in Fig. \ref{fig_8} with different SNRs. 
As we can see, the MSE performance of the proposed detector
converge after a certain number of iterations and the converge MSE performance benefits from higher SNR.
In addition, the state evolution of the proposed CD-D-OAMP detector can closely match to the actual MSE performance, even for large value of $C$, which indicates the effectiveness of our analysis for state evolution.

\subsection{Complexity Analysis}
\begin{table}[t!]
	\renewcommand{\arraystretch}{1.5}
	\begin{center}
		\caption{Complexity comparison of different detectors}
		\begin{tabular}{c|c}
			\hline 
			\text{Algorithm} & \text{Computation Complexity} \\ 
			\hline 
			CD-D-OAMP [Pro.] & \!\! $ \!\! \mathcal{O} \left(n_{t} \!\!\left(\! CN_c^3\!+\!CN_c^2 |\mathcal{D}_c|\!+\! 2N\log N \!+\!  4 N \!+\! N Q \!\right)\! \right) $  \\ 
			D-OAMP [\textbf{Re. 5}]  & $\mathcal{O} \left( n_t \left(  C N_c^3 + C N_c^2 |\mathcal{\tilde{D}}_c| + N Q \right) \right) $  \\
			OAMP \cite{d31} & $\mathcal{O} \left( n_t \left(  2 N^3 + N Q \right) \right)  $  \\
			AMP \cite{d29} & $\mathcal{O} \left( n_t \left(   N |\mathcal{S} | + N Q \right) \right)  $ \\
			GMP \cite{d24} & $\mathcal{O} \left( n_t \left(  N  Q  |\mathcal{S} |\right) \right)  $  \\
			LMMSE \cite{d22} & $\mathcal{O} \left(  2 N^3 +  NQ \right)  $  \\
			\hline
		\end{tabular}
	\end{center}
\end{table}

From the \textbf{Algorithm 1} discussion, the proposed CD-D-OAMP detector is composed of multiple modules. This allows us to analyze the detection complexity of each module and determine the total complexity for each iteration. 
It is obvious that the complexity of the D-LMMSE estimator is from the $C$ group matrix-by-matrix products with complexity $\mathcal{O} (C N_c^2 |\mathcal{D}_c|)$ and matrix inverse with complexity $\mathcal{O} (C N_c^3)$. Typically, the $|\mathcal{D}_c|$ and $N_c$ is much smaller than $N$. Thus, the overall complexity order of the D-LMMSE estimator can be given by $\mathcal{O} \left( CN_c^3 + C N_c^2 |\mathcal{D}_c|\right) $. 
The complexity of cross domain operations are mainly from the DAF transform matrix $\mathbf{U}$ and IDAF transform matrix $\mathbf{U}^{\text{H}}$. Note that the transform matrix $\mathbf{U}$ is consisted of diagonal matrices $\boldsymbol{\Lambda}_{c_1}, \boldsymbol{\Lambda}_{c_2}$ and $N$-point DFT matrix $\mathbf{F}_N$, where the complexity of multiplying a diagonal matrix with a vector is $\mathcal{O} (N)$. The DFT can be efficiently implemented by fast Fourier transform (FFT) with a computational complexity of $\mathcal{O} (N \log N)$. Therefore, the complexity order of two cross domain transform operations is $\mathcal{O} (2 N \log N + 4 N)$. The DAF domain detection only involves the component-wise operation of all detection symbols and constellations, with a complexity $\mathcal{O} (N Q)$. Finally, the total complexity of the proposed CD-D-OAMP detector is given by $\mathcal{O} \left( C N_c^3 + C N_c^2 |\mathcal{D}_c| + 2 N \log N + 4 N + N Q \right) $ for each iteration. 

\begin{figure}[t!]
	\centering
	\includegraphics[scale=0.55]{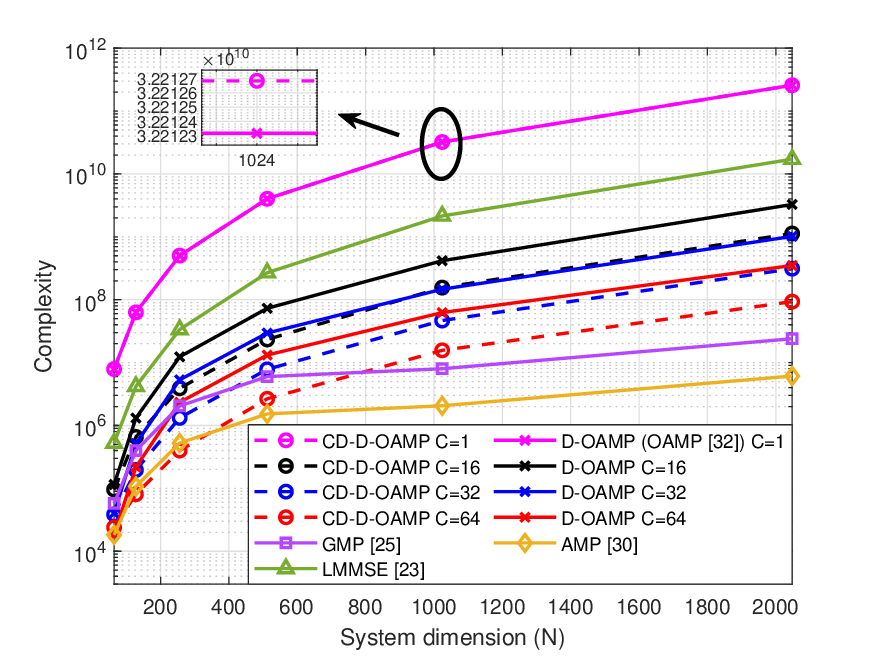}
	\caption{Complexity comparison of different detectors.}
	\label{fig_9}
\end{figure}

We summarize and compare the complexity of the proposed CD-D-OAMP detector and other existing detectors in Table \uppercase\expandafter{\romannumeral1}, where $n_{t}$ is the number of iterations, $\mathcal{\tilde{D}}_c$ is the index set of DAF domain signal $\mathbf{x}$ linked to correspondingly $c$-th group received signal $\mathbf{y}_{c}$ with size $|\mathcal{\tilde{D}}_c| $, $\mathcal{S}$ is the index set according to non-zero elements for each row of DAF domain channel matrix $\bar{\mathbf{H}}$ with size $|\mathcal{S}|$. 
To intuitively compare the algorithms complexity of different detectors, we show the variation trend of computational complexity versus system dimension $N$ in Fig. \ref{fig_9} and set $n_t = 15$.
It can be observed that the complexity of the proposed CD-D-OAMP detector significantly decreases as $C$ increases due to smaller-dimensional matrices are processed by the LMMSE estimator.
It is also notice that the complexity of the proposed CD-D-OAMP is lower than that of D-OAMP for the same $C$, 
since the time domain equivalent channel matrix is more sparse than the DAF domain.
The slight complexity increasing of CD-D-OAMP detector compared to D-OAMP detector with $C=1$ is primarily from the cross domain operations. The LMMSE detector \cite{d22} shows high complexity due to matrix inversion. Although the GMP \cite{d24} and AMP \cite{d29} detectors exhibit 
particularly  low complexity advantageous for large-dimensional AFDM systems, they will suffer severe performance loss due to the DAF domain equivalent channel sparsity reduction under time-scaled wideband doubly-dispersive channels, which will be discussed in Section \uppercase\expandafter{\romannumeral6}.
In addition, the proposed CD-D-OAMP detector can support the parallel computing processes, which can significantly reduce the computational latency.

\section{Simulation Results}
In this section, we evaluate the BER performance of the proposed CD-D-OAMP detector for AFDM system under time-scaled wideband doubly-dispersive channels. Without loss of generality, we consider the time-scaled wideband underwater acoustic channels and wideband wireless communication channels, respectively. The QPSK modulation is applied to AFDM system with the number of transmitted symbols $N$. To compare the performance of OTFS modulation, we set $N = \tilde{M} \tilde{N}$ and OTFS occupies the same time-frequency resources with AFDM and other modulation schemes, where $\tilde{M}$ and $\tilde{N}$ are the delay and Doppler dimensions in OTFS systems. In the time-scaled wideband underwater acoustic communication, we consider the number of transmitted symbols $N = 1024$ for OFDM, OCDM and AFDM systems, $\tilde{M} = 64$ and $\tilde{N} = 16$ for OTFS system.
The carrier frequency is $f_c = 6$kHz with sub-carriers spacing $\varDelta f = 4$Hz \cite{d41}. The maximum Doppler scale factor is $\alpha_{max} = 10^{-4}$. In the time-scaled  wideband wireless communication, we consider the Terahertz (THz) communication with $N = 65536$ for OFDM, OCDM and AFDM systems, $\tilde{M} = 1024$ and $\tilde{N} = 64$ for OTFS system. The carrier frequency is $f_c = 15$THz with sub-carriers spacing $\varDelta f = 2$MHz \cite{d44}. The maximum mobile velocity is $v_{max} = 500$km/h, leading to a maximum Doppler scale factor $\alpha_{max} = 4.6 \cdot 10^{-7}$.  Then
we generate the Doppler scale factor for each path as $\alpha_i = \alpha_{max} \cos (\tilde{\theta}_i)$, where $\tilde{\theta}_i$ is uniformly distributed over $[-\pi,\pi]$. We consider the number of channel paths $P = 4$ and the channel coefficients $h_i$ are randomly generated based on a uniform power delay profile. The maximum number of iterations is $n_{t} = 15$.

\begin{figure}[t!]
	\centering
	\includegraphics[scale=0.55]{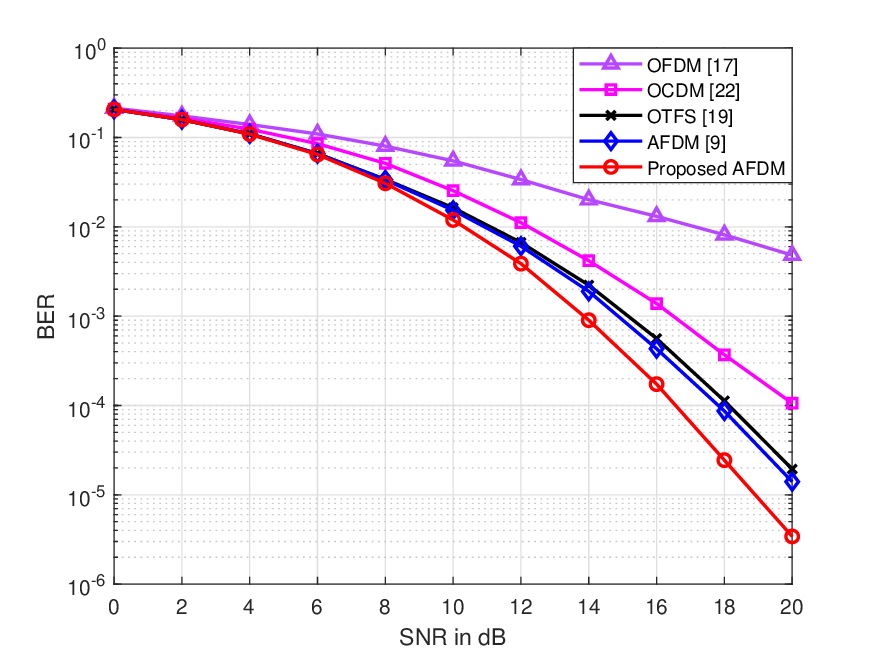}
	\caption{BER performance analysis with different modulation schemes under underwater acoustic channel, maximum Doppler scale factor $\alpha_{max} = 10^{-4}$.}
	\label{fig_10}
\end{figure} 

\begin{figure}[t!]
	\centering
	\includegraphics[scale=0.55]{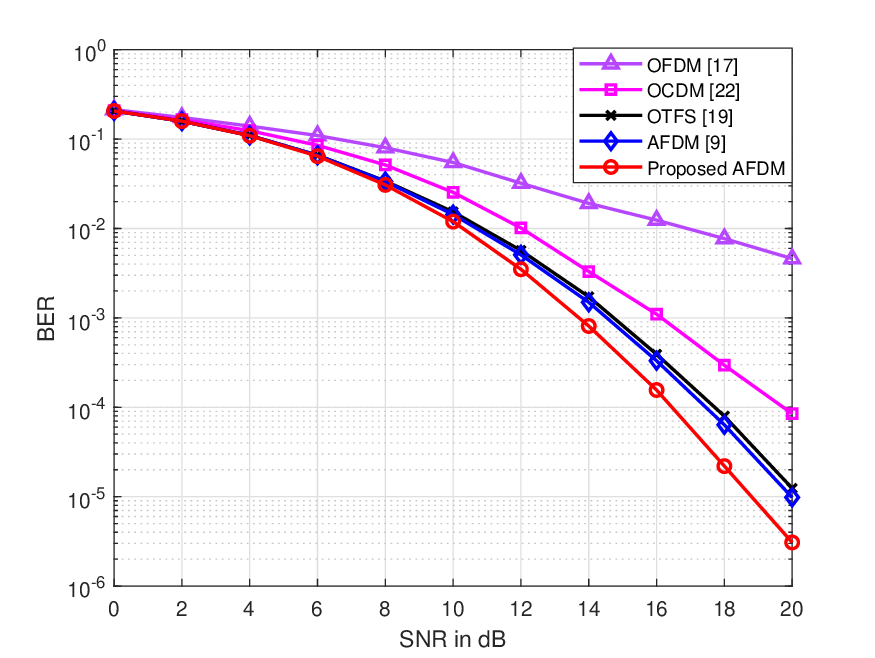}
	\caption{BER performance analysis with different modulation schemes under wireless communication channel, maximum Doppler scale factor $\alpha_{max} = 4.6 \cdot 10^{-7}$.}
	\label{fig_10_1}
\end{figure} 

We first consider a more practice large dimension system and compare the BER performance of different modulation schemes based on OAMP detection \cite{d31} under time-scaled wideband underwater acoustic channels and wideband wireless communication channels, as shown in Fig. \ref{fig_10} and Fig. \ref{fig_10_1}, respectively. The results demonstrate that our optimized wideband chirp parameters AFDM system achieves superior BER performance compared to OFDM \cite{d16}, OCDM \cite{d21}, OTFS \cite{d18} and AFDM with conventional narrowband chirp parameters \cite{d9}. These large dimension system results also align with the small dimension system observations in Fig. \ref{fig_4}. Furthermore, the proposed AFDM system maintains consistent performance advantages across both wideband underwater acoustic and wideband wireless communication environments. These comprehensive results robustly support the effectiveness of the AFDM system with our optimized wideband chirp parameters in time-scaled wideband doubly-dispersive channels. Since the performance of AFDM system  exhibits similar results in time-scaled wideband underwater acoustic channels and wideband wireless communication channels observed from Fig. \ref{fig_10} and Fig. \ref{fig_10_1}, we take the time-scaled wideband underwater acoustic channel as an example in subsequent simulations to evaluate the BER performance of the proposed detectors. Note that the results can also be extended to the time-scaled wideband wireless communication scenarios.

\begin{figure}[t!]
	\centering
	\includegraphics[scale=0.55]{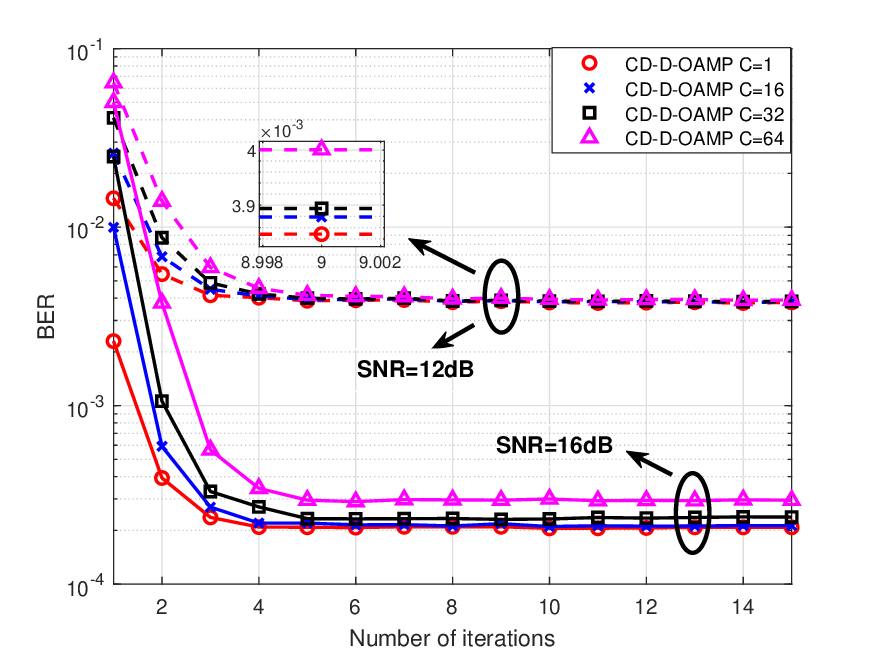}
	\caption{Convergence analysis with different iteration numbers.}
	\label{fig_11}
\end{figure} 

We then analyze the convergence behavior of the proposed CD-D-OAMP detector with different $C$, as shown in Fig. \ref{fig_11}. It is observed that our proposed CD-D-OAMP detector converges after a certain number of iterations and achieves better BER performance benefits from higher SNR. Although the convergence rate decreases slightly as $C$ increases, it has a significantly complexity reduction observed from Fig. \ref{fig_9}. 
More importantly, for a sufficiently large $C$, there is only a slight performance loss compared to the CD-D-OAMP detector with $C=1$ but exhibits significantly  complexity reduction.

\begin{figure}[t!]
	\centering
	\includegraphics[scale=0.55]{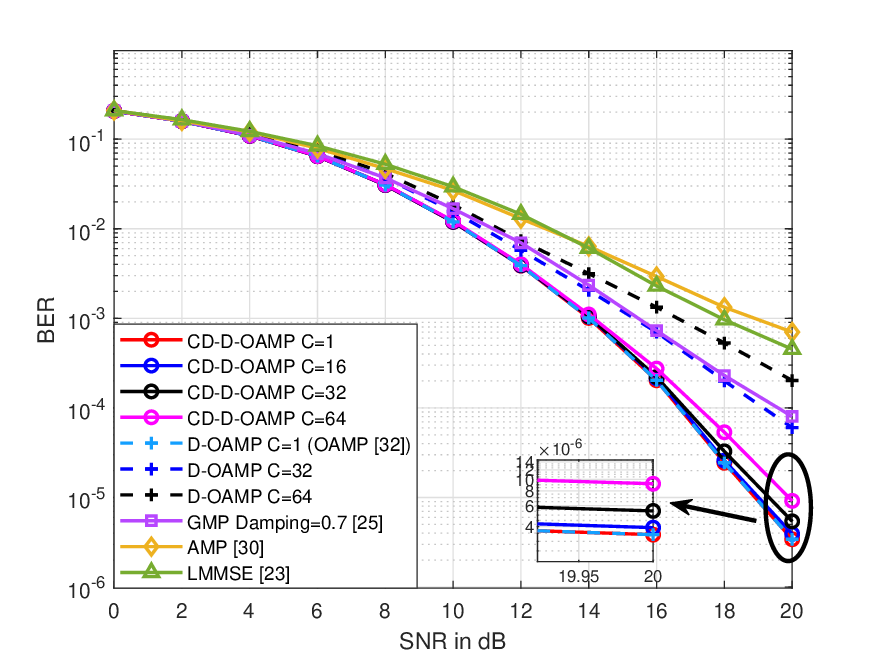}
	\caption{BER performance analysis with different detectors.}
	\label{fig_12}
\end{figure} 

Finally, we compare the BER performance of AFDM system with different detectors under time-scaled wideband doubly-dispersive channels in Fig. \ref{fig_12}.
We can observe that the OAMP detector \cite{d31} shows the better BER performance, but the high computational complexity, which limits its application in practice. The LMMSE detector \cite{d22} exhibits a worse performance due to ignoring the nonlinear prior distribution  constraints. The low-complexity detectors such as GMP \cite{d24} and AMP \cite{d29} do not show the expected performance due to the large number of four-edge cycles in the corresponding factor graph and non-i.i.d Gaussian channel matrix.
However, our proposed CD-D-OAMP detector can achieve the better detection performance under time-scaled wideband doubly-dispersive channels. 
Even with sufficiently large value of $C$ such as  $C=64$, there is only a slight performance loss but significant complexity reduction. In addition, our proposed CD-D-OAMP detector shows stronger robustness than D-OAMP detector with same $C$. This benefits from the lower interference in sparser time domain equivalent channel, which enables the more accurate estimation of D-LMMSE estimator. Therefore, our proposed CD-D-OAMP detector can achieve 
desirable trade-off between complexity
and performance.


\section{Conclusion}
In this paper, we investigated the AFDM systems in wideband doubly-dispersive channels with time-scaling effects. Firstly, we developed an efficient transmission structure with CPP and CPS for AFDM system and derived the corresponding input-output relationship under time-scaled wideband doubly-dispersive channels. 
We further optimized the AFDM chirp parameters and demonstrated the superiority of our optimized chirp parameters by PEP analysis.
We also proposed an efficient CD-D-OAMP algorithm for AFDM symbols detection and derived the correspondingly state evolution. Through a comprehensive analysis of the detection complexity and BER performance evaluation based on simulation, the AFDM systems with our optimized chirp parameters outperform the other modulation schemes in time-scaled wideband doubly-dispersive channels. Moreover, the proposed CD-D-OAMP detector can achieve the desirable trade-off between the complexity and performance, while significantly reducing the computational latency.

\vfill

\end{document}